\documentclass[aps,prd,preprint,superscriptaddress,nobibnotes,nofootinbib,a4paper]{revtex4-1}
\usepackage[english]{babel}
\usepackage{graphicx}
\usepackage{amsmath,amssymb}
\usepackage[caption=false]{subfig}
\usepackage{setspace}
\usepackage{hyperref}
\usepackage{rotating}

\newcommand{\p}{\partial}

\newcommand{\Tr}{\mathop{\rm Tr}\nolimits}

\newcommand{\beq}{\begin{eqnarray}}
\newcommand{\eeq}{\end{eqnarray}}
\newcommand{\non}{\nonumber\\}

\begin{document}

\title{Modifying the pion mass in the loosely bound Skyrme model}

\author{Sven Bjarke Gudnason}
\email{bjarke(at)impcas.ac.cn}
\affiliation{Institute of Modern Physics, Chinese Academy of Sciences,
  Lanzhou 730000, China}
\author{Muneto Nitta}
\email{nitta(at)phys-h.keio.ac.jp}
\affiliation{{\it Department of Physics, and Research and
    Education Center for Natural Sciences, Keio University, Hiyoshi
    4-1-1, Yokohama, Kanagawa 223-8521, Japan}}

\date{\today}
\begin{abstract}

We study the loosely bound Skyrme model with the addition of two
different pion mass terms; this is the most general potential of
polynomial form up to second order in the trace of the Skyrme field.
The two pion mass terms are called the standard pion mass term and the
modified pion mass term.
We find that the binding energies are not reduced by the introduction
of the modified pion mass, but it is analogous to the standard pion
mass term with a decrease in the value of the mass parameter of the
loosely bound potential (for large values of the latter parameter).
We find by increasing the overall pion mass that we can reduce the
classical binding energy of the 4-Skyrmion to the 2.7\% level and the
total binding energy including the contribution from spin/isospin
quantization is reduced to the 5.8\% level.

\end{abstract}


\maketitle

\section{Introduction}

The Skyrme model was made as an effective theory of pions that could
describe baryons in terms of its soliton -- the Skyrmion
\cite{Skyrme:1962vh,Skyrme:1961vq}.
It was, however, not taken too serious as a model until Witten pointed
out that the Skyrmion should be identified with the baryon in
large-$N_c$ quantum chromodynamics
\cite{Witten:1983tw,Witten:1983tx}.
Although the single and charge-two Skyrmions were studied in the
literature in the following years, little progress was made on finding
Skyrmions with higher baryon numbers (three and above) until the
idea of using rational maps was introduced
\cite{Battye:1997qq,Houghton:1997kg}.
The Skyrmion solutions were then soon found and their symmetries
identified for baryon numbers up to and including $B=22$
\cite{Battye:2001qn}.
These Skyrmions are well described by rational maps and look like
fullerenes and thus they are hollow, almost spherical shells of baryon
charge $B$ with $2B-2$ holes in them.
This approach seemed to be on the right track as it is a convenient
and precise way of finding Skyrmion solutions with higher baryon
numbers.
For the single Skyrmion the pion mass term has little qualitative
effect; naively it seems that it just decreases the size slightly and
increases the energy a little \cite{Adkins:1983hy}; nothing that
refitting the parameters cannot compensate.
For the Skyrmions of higher baryon numbers, however, it turned out that
the pion mass has a drastic effect; the fullerene-type hollow shells
are only the preferred minima of the energy when the pion mass is
turned off or very small
\cite{Battye:2004rw,Battye:2006tb,Houghton:2006ti}.
In fact, for a pion mass of the order of its experimentally measured
value, the Skyrmions prefer to order themselves as cubes in a crystal
-- akin towards the alpha-particle model of nuclei
\cite{Battye:2006na}.
However, the Skyrmions are much more complex than just point particles
with interactions and thus should not be directly compared to the 
alpha-particle model.

All these steps of progress towards finding Skyrmion solutions of
higher baryon numbers brought us to this point and in principle
Skyrmion solutions of any baryon number can now be constructed.
The Skyrme model as was used up to this point is made of three terms;
the kinetic term, the Skyrme term and the linear pion mass term
(linear in the chiral Lagrangian field $U$). We shall henceforth call
this pion mass term the \emph{standard} pion mass term.
However, a notorious problem has been tagging along so far; namely the
binding energies of the Skyrmions with higher baryon numbers are much
too large; they are about one order of magnitude larger than the
experimentally observed values. 
This problem motivated several directions of improving the standard
Skyrme model.
One attempt at mending the problem of the large binding energies was
the idea of starting from a higher-dimensional self-dual theory,
perform dimensional reduction and then identify the Skyrme model as
the leading order Lagrangian; the binding energy in this construction
would go to zero if infinitely many mesons were to be integrated in
\cite{Sutcliffe:2010et,Sutcliffe:2011ig}.
Another direction is based on the discovery of a subsector where the
model has a Bogomol'nyi bound that can actually be saturated
\cite{Adam:2010fg,Adam:2010ds}; unlike that of the standard Skyrme
model \cite{Manton:1986pz}.
This model is constructed by squaring the baryon charge current and
adding a potential and is by now called the BPS-Skyrme model. 
One peculiarity of this model is that it does not contain a kinetic
term and not the Skyrme term either.
A strength of this model is that it models a perfect fluid, which is a
welcomed feature in the light of nuclear matter and neutron stars
\cite{Adam:2014nba,Adam:2014dqa,Adam:2015lpa,Adam:2015lra}.
In a realistic model of nuclei, however, one would expect the presence
of at least the kinetic term in the model.
Turning on the kinetic term and the Skyrme term with order-one
coefficients, however, renders the model very similar to the standard
Skyrme model and the binding energies are again too large.
One idea is then that the kinetic term and the Skyrme term are rather
small compared to the BPS-Skyrme term \cite{Adam:2015ele}.
This turns out to be a rather difficult technical problem; what
happens here is that when only the BPS term is present in the theory
(plus a potential), then the Skyrmions can take any shape.
However, with the kinetic term and the Skyrme term turned on, the
Skyrmions like to take their usual shapes of platonic solids; however,
if the coefficient of the latter two terms is very small, then the
solutions can afford very large derivatives.
This fact is quite a problem for most codes for Skyrmion calculations 
\cite{Gillard:2015eia}.
The third direction of reducing the binding energies in the
Skyrme-like models, is to take the standard Skyrme model and add to it
a holomorphic (quartic) potential, which is based on an energy bound
that, however, can only be saturated for the single Skyrmion
\cite{Harland:2013rxa,Gillard:2015eia}.
This model was called the lightly bound Skyrme model in
Ref.~\cite{Gillard:2015eia}.
Although the lightly bound Skyrme model, i.e.~the Skyrme model with
the holomorphic potential, is able to reduce the binding energies of
the multi-Skyrmions; long before reaching experimentally observed
values, the symmetries of the Skyrmions completely change and the
platonic symmetries are lost \cite{Gudnason:2016mms}.
This leads to severe problems of retaining the earlier successes of
the Skyrme model; in particular if the cubic shape of the 4-Skyrmion
is lost, then the identification of the Hoyle state and the ratio of
slopes of the ground state and Hoyle state rotational bands
\cite{Lau:2014baa} should be reconsidered entirely.
A related problem with the lightly bound Skyrme model is that the
binding energy of the $B=5$ Skyrmion is higher than that of the $B=4$ 
Skyrmion and hence nuclear clustering \cite{Freer:2007} into $n$ alpha
particles for nucleon number $A=4n$, is no longer possible. 
In Ref.~\cite{Gudnason:2016mms} we have chosen to keep the cubic
symmetry of the 4-Skyrmion to retain the clustering of the nuclei;
which in our opinion is a strength of the Skyrme-type models.
Not only trying to keep the symmetries and hence the successes of the
Skyrme model, a better potential than that of the lightly bound Skyrme
model was found in Ref.~\cite{Gudnason:2016mms}; we call the Skyrme
model with this quadratic potential the loosely bound Skyrme model. 
The loosely bound Skyrme model can reach lower binding energies than
the lightly bound model before the symmetries change from platonic to
face-centered cubic (FCC) symmetries\footnote{Let us clarify that we
  use the term \emph{face-centered cubic (FCC)} in this paper to refer
to the Skyrmions that split up into separate $B=1$ clumps of baryon
charge situated at the vertices of a cubic lattice
\cite{Gillard:2015eia}. Obviously, for finite-sized Skyrmions it then
only corresponds to a part cut out from the lattice. In particular for
the 4-Skyrmion that we will study in this paper, the symmetry turns
from cubic to tetrahedral; we will nevertheless call the tetrahedral
state \emph{FCC}. }. 

As pointed out several times in the literature, the pion mass can be
made from infinitely many different terms, see
e.g.~\cite{Marleau:1990nh,Kopeliovich:2005vg,Piette:2008ch,Davies:2009zza,Gudnason:2016mms}.
In Ref.~\cite{Gudnason:2016mms} a class of potentials giving rise to a
pion mass term was contemplated
\beq
V_{0n} = \frac{1}{n} m_{0n}^2 \left(1 - \sigma^n\right),
\eeq
where $\sigma=\Tr[U]/2$ and $U$ is the Skyrme field related to the
pions as $U=\sigma\mathbf{1}_2+i\pi^a\tau^a$.
For each $n$ the above potential gives a normalized mass term for the
pions. Only the sum of these terms is measured.
The loosely bound potential, on the other hand, belongs to a class of
potentials that does not contribute to the pion mass
\beq
V_n = \frac{1}{n} m_n^2 \left(1 - \sigma\right)^n, \qquad n\geq 2.
\eeq
The loosely bound potential corresponds to $n=2$ and the lightly bound
potential corresponds to $n=4$.
Notice that the two classes of potentials coincide for $n=1$. 

Although the pion decay constant and the pion mass are both
experimentally known quantities, a modern point of view in the Skyrme
model is to consider them as renormalized (effective) constants, that
should be renormalized in the baryon medium and not in the pion
vacuum (i.e.~at zero chemical potential and zero temperature).
Therefore the pion decay constant is often taken to be around half of
its measured value.\footnote{Ref.~\cite{Battye:2005nx} also argues
  that the mass of the delta resonance and nucleon mass can only be
  fitted in the standard Skyrme model if the pion mass is taken to be
  larger than its measure value. }
In this spirit, we will in this paper also allow for some slush in the
pion mass and consider values too small and too large, in order to
study the effects on the model. 

In this paper, we will take the loosely bound Skyrme model, which is
the Skyrme model with the potential $V_2$ as well as the two first
terms contributing to the pion mass, i.e.~$V_{01}=V_1$ and $V_{02}$.
This is the most general potential of polynomial form up to second
order in $\sigma$.
Let us first contemplate what effect we could expect from switching
the standard pion mass term, $V_1$, with the modified pion mass term
$V_{02}$
\cite{Piette:1997ce,Kudryavtsev:1999zm}. 
Since the Skyrmion of charge $B$ needs to wrap a 3-cycle on the target
space $B$ times, it
will necessarily pass the antipodal point to the vacuum ($\sigma=-1$)
$B$ times. When the Skyrme field is near this antipodal point the
standard pion mass term has its maximal contribution to the energy,
whereas the modified pion mass term has none. 
Since the binding energy is a comparison between the 1-Skyrmion and
the $B$-Skyrmion, we can easily see that increasing the energy more
for the $B$-Skyrmions than for the 1-Skyrmion, lowers the relative
binding energy of the $B$-Skyrmion.
Therefore one would naively conclude that the standard pion mass term
is preferred over the modified one.

This paper is thus a complete scan of the parameter space of the most
general potential of polynomial form up to second order in $\sigma$.
We find in agreement with the above contemplation of the modified pion
mass that it increases the binding energy of the $B$-Skyrmions.
However, although the increase in binding energy is considerable when
the loosely bound potential is turned off, it becomes smaller and
almost insignificant when the coefficient of the loosely bound
potential is turned to its maximal value; i.e.~just before the cubic
symmetry of the 4-Skyrmion is lost.
This is related to the fact that in this region of parameter space,
switching from the linear or standard pion mass term to the modified
pion mass term merely results in a lower value of the mass parameter
of the loosely bound potential.
This can thus be compensated by increasing the value of the latter
mass parameter. 
We study the effects of the complete parameter space on the classical
binding energy, the total binding energy which takes into account the
quantum contribution from the spin and isospin of the nucleon, the
pion decay constant, the mass spectrum, and finally the charge radius
of the proton. 
We are able to reduce the classical binding energy to about the 2.7\% 
level and the total binding energy to about the 5.8\% level.
The conclusion is that the modified pion mass term is not
advantageous, but an increase in the value of the pion mass allows for
a larger value of the mass parameter of the loosely bound potential,
which in turn lowers the binding energy further. 

The paper is organized as follows.
In Sec.~\ref{sec:model} we present the loosely bound Skyrme model with
the two different pion mass terms; i.e.~the most general potential up
to second order in $\sigma$; set the notation and define the
observables that we will study on the entire parameter space of the
model. 
Sec.~\ref{sec:numerical} explains the numerical methods used and
Sec.~\ref{sec:results} presents the results.
Finally Sec.~\ref{sec:discussion} concludes with a discussion and
Appendix \ref{app:PT} shows figures of numerical Skyrmion solutions at
the boundary between the cubic and FCC symmetry regions in the
parameter space.

\section{The model and observables}\label{sec:model}

The model under study is the Skyrme model and the Lagrangian density
in physical units reads
\beq
\mathcal{L} = \frac{f_\pi^2}{4} \mathcal{L}_2
+ \frac{1}{e^2} \mathcal{L}_4
- \frac{\tilde{m}_\pi^2 f_\pi^2}{4 m_\pi^2} V,
\eeq
where the kinetic (Dirichlet) term and Skyrme term is given by
\begin{align}
  \mathcal{L}_2 = \frac{1}{4}\Tr(L_\mu L^\mu), \qquad
  \mathcal{L}_4 = \frac{1}{32}\Tr\left([L_\mu,L_\nu][L^\mu,L^\nu]\right),
\end{align}
and $L_\mu \equiv U^\dag \p_\mu U$.
$f_\pi$ is the pion decay constant with units of energy (MeV), $e>0$
is a real-valued dimensionless constant, $\tilde{m}_\pi$ is the pion
mass in MeV and, finally, $m_\pi$ is a dimensionless pion mass
parameter. 
The indices $\mu,\nu=0,1,2,3$ are spacetime indices and $U$ is the 
Skyrme field which in terms of the pions reads 
\beq
U = \mathbf{1}_2\sigma + i\tau^a\pi^a,
\eeq
with $U^\dag U=\mathbf{1}_2$ being the nonlinear sigma model
constraint, which is equivalent to $\sigma^2+\pi^a\pi^a=1$ and
$\tau^a$ are the Pauli matrices. 

It will prove convenient to do a rescaling of the energy and length
scales and only work with dimensionless parameters.
In particular, we will make a rescaling such that
$\tilde{x}^i=\mu x^i$, where both $\tilde{x}^i$ and $\mu$ have units
of length (MeV${}^{-1}$), and similarly for the energy
$\tilde{E}=\lambda E$; where $\tilde{E}$ and $\lambda$ have units of
energy (MeV). In particular, we get
\beq
\mathcal{L} = c_2 \mathcal{L}_2 + c_4 \mathcal{L}_4 - V,
\label{eq:L}
\eeq
where $c_2>0$ and $c_4>0$ are positive-definite real constants and
\beq
\lambda = \frac{f_\pi}{2e\sqrt{c_2c_4}}, \qquad
\mu = \sqrt{\frac{c_2}{c_4}} \frac{2}{e f_\pi},
\label{eq:unit_conversion}
\eeq
whereas the pion mass in physical units (MeV) is given by
\beq
\tilde{m}_\pi = \frac{\sqrt{c_4}}{2c_2} e f_\pi m_\pi.
\label{eq:mpion_physical_units}
\eeq
This relation assumes that the potential will have a pion mass
normalized to $m_\pi$ in dimensionless units. 

The main focus of this paper is to study the most general potential of
polynomial form up to second order in $\Tr[U]$: 
\beq
V = V_1 + V_{02} + V_2,
\label{eq:V}
\eeq
and the potentials are defined as
\begin{align}
  V_1 \equiv m_1^2 (1 - \sigma), \qquad
  V_{02} \equiv \frac{1}{2} m_{02}^2 (1 - \sigma^2),\qquad
  V_2 \equiv \frac{1}{2} m_2^2 (1 - \sigma)^2,
\end{align}
where the mass parameters $m_1, m_{02}, m_2$ are all real and 
\beq
\sigma = \frac{1}{2}\Tr[U].
\eeq
Note that there are only 2 free parameters to second order because the
constant is irrelevant for the equations of motion.
However, the above basis is convenient because all mass parameters are
real-valued.
We will nevertheless change to a simpler basis shortly. 

The Lagrangian density \eqref{eq:L} without a potential turned on,
enjoys SU(2)$\times$SU(2) symmetry. This symmetry is explicitly broken
down to a diagonal SU(2) by the potential \eqref{eq:V}. This SU(2)
corresponds to isospin and we will keep it unbroken in this paper. 

The target space of the Skyrme model -- due to the mentioned symmetry
breaking -- is given by $\mathcal{M}\simeq\mathrm{SU}(2)\simeq S^3$.
The map $U$ -- the Skyrme field -- is thus a map from space,
i.e.~$\mathbb{R}^3\cup\{\infty\}\simeq S^3$ to the target space
$\mathcal{M}$ and is characterized by the third homotopy group 
\beq
\pi_3(\mathcal{M}) = \mathbb{Z} \ni B,
\eeq
which admits solitons called Skyrmions and the integer $B$ is called
the baryon number, which in turn can be calculated from the baryon
charge density
\beq
\mathcal{B}^0 = -\frac{1}{12}\epsilon^{ijk} \Tr[L_i L_j L_k],
\label{eq:Bdensity}
\eeq
by integrating over space
\beq
B = \frac{1}{2\pi^2}\int d^3x\; \mathcal{B}^{0}.
\eeq

The pion mass (squared) in the model is given by
\beq
m_\pi^2  = - \left.\frac{\p V}{\p\sigma}\right|_{\sigma=1},
\eeq
and as explained in Ref.~\cite{Gudnason:2016mms}, the potentials $V_1$ 
and $V_{02}$ belong to the class of potentials giving rise to a pion
mass in the vacuum $\sigma=1$, whereas $V_2$ gives no contribution to 
the pion mass.
In particular, calculating the pion mass from the potential
\eqref{eq:V}, we get
\beq
m_\pi^2 = m_1^2 + m_{02}^2,
\eeq
and so we can parametrize the two mass parameters giving a pion mass
contribution as 
\beq
m_1^2 = \alpha m_\pi^2, \qquad
m_{02}^2 = (1-\alpha) m_\pi^2,
\eeq
where the real parameter $\alpha\in[0,1]$ takes on a value in the
interval from zero to one; $\alpha=1$ corresponds to the traditional
pion mass, whereas $\alpha=0$ yields the modified pion mass
\cite{Piette:1997ce,Kudryavtsev:1999zm} 
and any value in between is a linear interpolation between the two.

We will now switch to a simpler basis for the potential
\begin{align}
  V &= \alpha m_\pi^2(1 - \sigma)
  + \frac{1}{2}(1 - \alpha) m_\pi^2(1 - \sigma^2)
  + \frac{1}{2}m_2^2(1 - \sigma)^2 \non
  &= m_\pi^2(1 - \sigma)
  + \frac{1}{2}\left[m_2^2 - (1-\alpha)m_\pi^2\right] (1 - \sigma)^2,
\end{align}
where we have absorbed the modified pion mass term into the loosely
bound potential in the second line.
If we set $\alpha=1$ then the potential is equal to a subset of that
analyzed in Ref.~\cite{Gudnason:2016mms}.
However, when $\alpha<1$ the coefficient of $(1-\sigma)^2$ (the
loosely bound potential term) is no longer positive semi-definite;
i.e.~the mass parameter is no longer only real-valued.
We will now define the parameter
\beq
\mathbf{m}_2^2 \equiv m_2^2 - (1-\alpha)m_\pi^2,
\label{eq:boldm2sq}
\eeq
which takes values in the range $[-m_\pi^2,\infty)$ and the potential
is then simply
\beq
V = m_\pi^2 (1 - \sigma)
+ \frac{1}{2} \mathbf{m}_2^2 (1 - \sigma)^2.
\label{eq:Vsimp}
\eeq
It is easy to confirm that $\sigma=1$ is always a (local) vacuum.
The lower bound on $\mathbf{m_2}^2$ comes from the condition that
$\sigma=-1$ should not become the global vacuum; the value of
$\mathbf{m}_2^2$ where the two vacua, $\sigma=\pm 1$ become degenerate
is exactly $\mathbf{m}_2^2=-m_\pi^2$.
There is no upper bound on $\mathbf{m}_2^2$, however, when the
parameter becomes too large, the platonic symmetries of the
multi-Skyrmions are lost; in particular, the 4-Skyrmion loses its
cubic symmetry \cite{Gudnason:2016mms}.

Now we can see from Eq.~\eqref{eq:boldm2sq} that when $\alpha=1$,
$\mathbf{m}_2=m_2$ as expected and when $\alpha<1$, the modified pion
mass term is turned on, corresponding to a decrease in the effective
value of the mass parameter $m_2$.
In order to cover the complete parameter space, however, we need to
consider also the negative range of $\mathbf{m}_2^2$, corresponding to
the case where $m_2^2<(1-\alpha)m_\pi^2$, which is possible only when
the modified pion mass is turned on.
When $m_2\gg m_\pi$, we can thus expect that the modified pion mass
does not provide any advantage at all, since it merely reduces the
effective value of $m_2$ and from Ref.~\cite{Gudnason:2016mms} we know
that the largest possible value of $m_2$ provides the lowest possible
binding energies in that range.
The negative range of $\mathbf{m}_2^2\in[-m_\pi^2,0)$ is, however,
until now unexplored.

Now we should make a choice concerning the (dimensionless) units,
i.e.~fixing $c_2$ and $c_4$.
The standard Skyrme units correspond to $c_2=c_4=2$ for which the
energy and length are given in units of $f_\pi/(4e)$ and $2/(ef_\pi)$,
respectively, see Ref.~\cite{MantonSutcliffe}. 
Here we will apply the same convention for units as used in
Ref.~\cite{Gudnason:2016mms}, namely
\beq
c_2 = \frac{1}{4}, \qquad
c_4 = 1,
\label{eq:c2c4}
\eeq
and hence energies and lengths according to
Eq.~\eqref{eq:unit_conversion} will be given in units of $f_\pi/e$ and
$1/(ef_\pi)$, respectively.  
The pion mass in physical units \eqref{eq:mpion_physical_units} with
the normalization convention \eqref{eq:c2c4} thus reads 
\beq
\tilde{m}_\pi = 2e f_\pi m_\pi.
\label{eq:physical_mp}
\eeq
Due to the different normalization of the potential \eqref{eq:V} (and
of the Lagrangian density \eqref{eq:L}) by a factor of two, the pion
mass $m=1$, used in Ref.~\cite{Battye:2006na}, corresponds to
$m_\pi=1/4$ and $\tilde{m}_\pi=e f_\pi/2$ in our units and
normalization.

Let us define the observables that we want to compare with data for
nuclei.
The first and the one of prime interest in this paper, is the
classical binding energy
\beq
\Delta_B \equiv B E_1 - E_B,
\eeq
where $E_B$ is the total energy of the Skyrmion with baryon number
$B$. 
It will however prove convenient to use the \emph{relative} classical
binding energy instead
\beq
\delta_B \equiv \frac{\Delta_B}{B E_1} = 1 - \frac{E_B}{B E_1},
\label{eq:deltaB}
\eeq
as the physical units drop out and we can use any units we like; in
particular our Skyrme units; i.e.~Skyrme units in our
normalization.
Before we can compare honestly with experiment, we should take into
account the quantum contribution to the ground state of spin and
isospin quantization, yielding
\beq
\delta_B^{\rm tot} \equiv
1 - \frac{E_B + \epsilon_B}{B(E_1 + \epsilon_1)},
\eeq
where $\epsilon_B$ is the quantum contribution to the ground state of
the baryon represented by the Skyrmion of baryon number $B$. 
Note in particular that the quantum contribution to the Skyrmion of
baryon number $B$ \emph{decreases} the binding energy, whereas the
quantum contribution to the 1-Skyrmion \emph{increases} the binding
energy.
As well known the quantum contribution to the 1-Skyrmion happens to be
larger than those to the higher $B$-Skyrmions and therefore the
spin-isospin quantization has the effect of increasing the already too
large binding energies of the Skyrmions.
The reason why the quantum contribution to the 1-Skyrmion is larger
than to the other ones is simply that the 1-Skyrmion is the smallest
one and hence it has the smallest moment of inertia.
Since the quantum contribution to the energy of the Skyrmion is
inversely proportional to the moment of inertia the above mentioned
effect on the binding energy follows. 

In this paper, for practical reasons of flops economy and because of
the fact that the ground state of the ${}^4$He is a spin-0, isospin-0
state, we will focus on the $B=1$ and $B=4$ sectors of the model; for
other baryon numbers in a subset of the model (the $\alpha=1$ sector),
see Ref.~\cite{Gudnason:2016mms}.
In particular, the latter fact implies that there is no quantum
contribution to the $B=4$ Skyrmion \cite{Manko:2007pr} and hence the
total relative binding energy simplifies to
\beq
\delta_4^{\rm tot} \equiv 1 - \frac{E_4}{4(E_1 + \epsilon_1)}.
\label{eq:delta4tot}
\eeq

Since we only need to calculate the ground state energy contribution
from spin-isospin quantization of the single Skyrmion, the calculation
is considerably simpler and we follow Ref.~\cite{Adkins:1983ya}.
Let $U\to A U A^{-1}$, with $A=A(t)$ being an SU(2) matrix which
rotates the isospin of $U$. This gives rise to the kinetic energy 
\beq
T = \frac{1}{2}a_i U_{ij} a_j = \Lambda \Tr(\p_0 A\p_0 A^{-1}),
\eeq
for the single ($B=1$) Skyrmion
\beq
U = \cos f(r) + i\hat{x}^i\tau^i \sin f(r),
\label{eq:hedgehog}
\eeq
where $a_i\equiv -i\,\Tr(\tau_i A^{-1}\dot{A})$, $\hat{x}^i$ is the
spatial unit vector and $U_{ij}=\Lambda\delta_{ij}$, with
\beq
\Lambda\equiv \frac{8\pi}{3}\int dr\; r^2\sin^2 f
\left(c_2 + c_4f_r^2 + \frac{c_4}{r^2}\sin^2 f\right),
\eeq
where $f_r\equiv \p_r f$.
Hence we get the kinetic energy from canonical quantization
\beq
T = \frac{1}{8\Lambda}\ell(\ell+2) = \frac{1}{2\Lambda}J(J+1).
\eeq
In particular, for the spin-1/2 ground state of the proton or neutron,
we get
\beq
T_{1/2} = \frac{1}{2\Lambda}\frac{3}{4},
\eeq
where $J=\ell/2$ is the spin quantum number.
Reinstating physical units and using our normalization
\eqref{eq:c2c4}, we get for the total mass 
\beq
\tilde{E}_1 + \tilde{\epsilon_1} =
\frac{f_\pi}{e} E_1 + \frac{e^3 f_\pi}{2\Lambda}\frac{3}{4}.
\eeq
When calculating the relative binding energy, the factor of $f_\pi/e$
will drop out, and hence we need only calculate
\beq
E_1 + \epsilon_1 =
\frac{e}{f_\pi}\left(\tilde{E}_1 + \tilde{\epsilon}_1\right)
= E_1 + \frac{e^4}{2\Lambda}\frac{3}{4}.
\eeq
We will also calculate the mass of the delta resonance. Since it is
merely the spin-3/2 state of the baryon in our model, we can estimate
the mass by
\beq
\tilde{m}_\Delta = \tilde{E}_1 + 5\tilde{\epsilon}_1.
\eeq

In order to add the quantum contribution to the energy of the
1-Skyrmion to its classical contribution, we need to fit the mass and
size of a selected Skyrmion to those of a corresponding nucleus.
A large part of the literature used the proton and delta resonance as
the two input parameters to fit $f_\pi$ and $e$ \cite{Adkins:1983ya};
this fit suffers from the problem that the binding energies for the
Skyrmions are about an order of magnitude too large compared with
those of nuclei. 
Later a different fit was made using ${}^6$Li \cite{Manton:2006tq};
the purpose here is to better match the energies of multi-Skyrmions
with higher $B$.
Other fits in the literature uses ${}^{12}$C \cite{Lau:2014baa} or
${}^4$He \cite{Gudnason:2016mms}.
For concreteness and simplicity, we will again use ${}^4$He as in
Ref.~\cite{Gudnason:2016mms}; this is convenient because we can use
the different 4-Skyrmions to do the calibration; in fact, we will
-- in this paper -- recalibrate each point in the parameter space of
the model such that the 4-Skyrmion fits the mass and size of
${}^4$He.
This will in turn give an accurate estimate of the effect on the
quantum contribution of the different parts of the parameter space. 

Another observable that we will calculate on the Skyrmion solutions is
the size of the nuclei. Since we fit the size of the 4-Skyrmion, we
will use that of the 1-Skyrmion as a check.
We choose to define the squared radius in terms of the baryon charge
density, i.e., 
\beq
r_B^2 = \frac{1}{2\pi^2 B}\int d^3x\; r^2\mathcal{B}^0,
\label{eq:rsqBdensity}
\eeq
where $\mathcal{B}^0$ is the baryon charge density given in
Eq.~\eqref{eq:Bdensity}.
Hence the size can be estimated as $r_B\sim\sqrt{r_B^2}$.

We are now ready to perform numerical calculations on the full
parameter space of the most general potential of polynomial form up to
second order in $\sigma$.

\section{Numerical calculations}\label{sec:numerical}

We will follow the approach used in Ref.~\cite{Gudnason:2016mms} and
use a finite difference method in conjunction with the relaxation
method for the partial differential equations (PDEs).
Our grid sizes are typically $101^3$ and we use a fourth-order
stencil.

In order to save computational costs, we use the hedgehog Ansatz
\eqref{eq:hedgehog} for calculating the 1-Skyrmions and solve the
ordinary differential equation (ODE)
\begin{align}
c_2\left(f_{rr} + \frac{2}{r}f_r - \frac{\sin 2f}{r^2}\right)
+c_4\left(\frac{2\sin^2(f)f_{rr}}{r^2} + \frac{\sin(2f)f_r^2}{r^2}
-\frac{\sin 2f\sin^2f}{r^4}\right)\non
= m_1^2\sin f + \frac{1}{2} m_{02}^2\sin 2f + m_2^2(1 - \cos f)\sin f,
\end{align}
to very high accuracy level; better than the $10^{-6}$ level.
Therefore, in order to compare the $B=4$ solutions to the $B=1$
solutions -- for the purpose of calculating the relative binding
energy -- we need to obtain the energy for the 4-Skyrmion very
precisely.
We will again utilize the trick used in Ref.~\cite{Gudnason:2016mms},
namely, we relax the numerical solution down to the $10^{-3}$ level,
locally (we denote this time $\tau_0$), and from then on, we make an 
exponential fit to the energy as function of relaxation (imaginary)
time. 
We continue the cooling process until the exponential fit is precise
enough and the imaginary time where we stop the calculation is
$\tau_2$. $\tau_1$ is defined as the midpoint:
$\tau_1=(\tau_0+\tau_2)/2$.
After the fit has been calculated, we take the $\tau\to\infty$ limit
of the energy function and the result is 
\beq
E_B\simeq \frac{B}{B_{\rm numerical}}\times
  \frac{E_{B,{\rm numerical}}(\tau_0)E_{B,{\rm numerical}}(\tau_2) -
    E_{B,{\rm numerical}}^2(\tau_1)}
  {E_{B,{\rm numerical}}(\tau_0) - 2E_{B,{\rm numerical}}(\tau_1) +
    E_{B,{\rm numerical}}(\tau_2)}.
\eeq
Note that we also use another trick of compensating the total energy
by a factor of $B/B_{\rm numerical}$ as both the energy and the baryon
charge is underestimated in the numerical calculation.
We have checked in Ref.~\cite{Gudnason:2016mms} that this reproduces
the energy for the $B=1$ sector within an accuracy of about
$2.7\times 10^{-4}$ or better.

As another check on the precision of the Skyrmion solutions, we
calculate the baryon charge numerically and find that all the
solutions yield $B=4$ to a precision of about $1.7\times 10^{-3}$ or
better. 
Therefore in summary our results should be trustable down to about the
permille level. 

For each data point in the parameter space, we refit the length and
energy scales to the ${}^4$He nucleus, thus determining $f_\pi$ and
$e$.
After the physical units are fitted we calculate all the observables
presented in the last section. 

We are now ready to present the results in the next section.

\section{Results}\label{sec:results}

\subsection{Classical binding energies}

\begin{figure}[!htp]
  \begin{center}
    \includegraphics[width=0.75\linewidth]{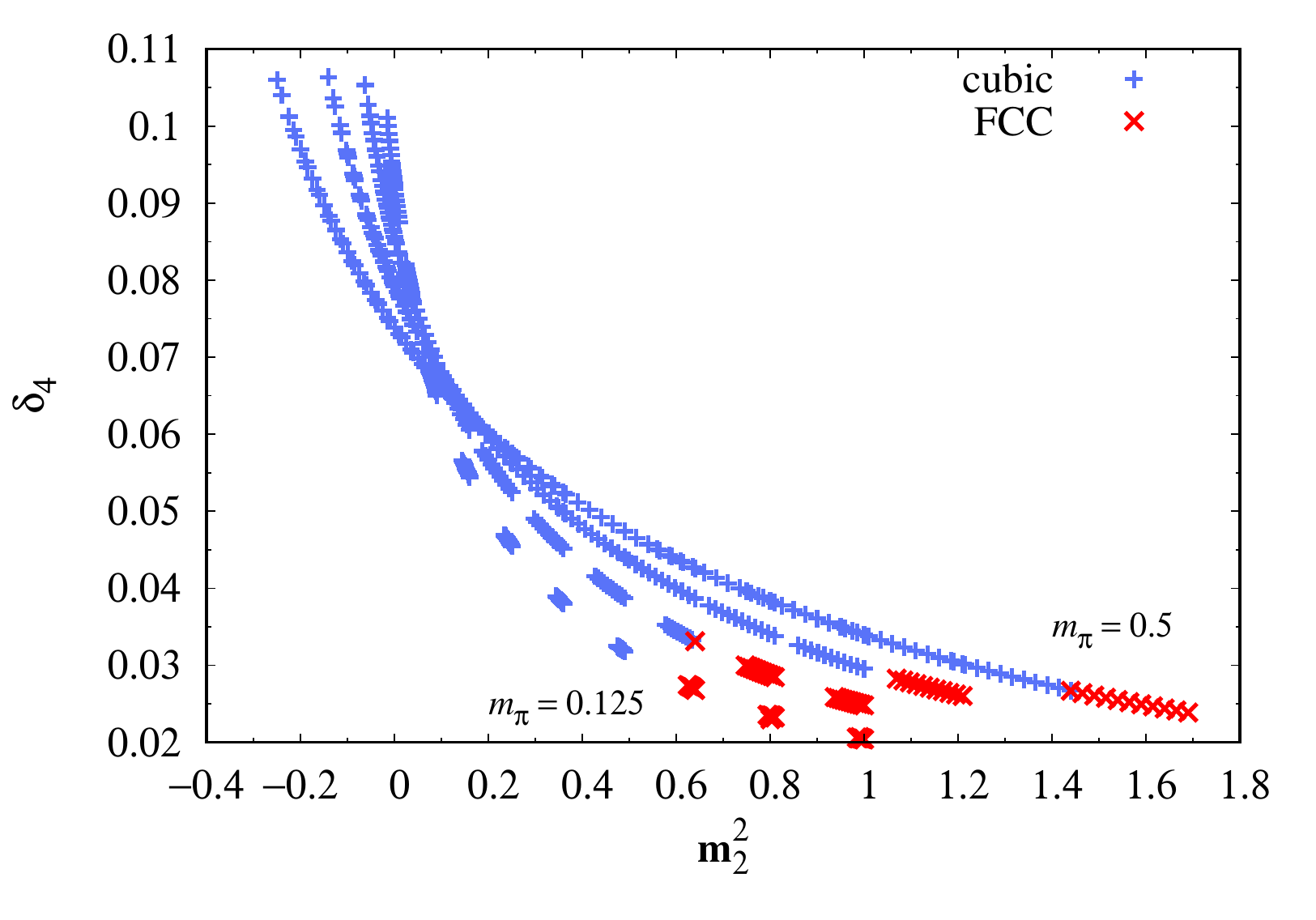}
    \caption{Classical binding energy $\delta_4$ as function of
      $\mathbf{m}_2^2$; four series of points are shown corresponding
      to $m_\pi=0.125,0.25,0.375,0.5$. }
    \label{fig:cbe}
  \end{center}
\end{figure}

We start by presenting the classical binding energies, defined in
Eq.~\eqref{eq:deltaB}, in Fig.~\ref{fig:cbe} for various values of
$m_\pi$ as functions of $\mathbf{m}_2^2$.
The four curves correspond to $m_\pi=0.125,0.25,0.375,0.5$.
The value $m_\pi=0.25$ corresponds to the choice $m=1$ in
Refs.~\cite{Battye:2006na,Manko:2007pr}. 
The blue dots in this figure and in the remainder of the paper,
correspond to Skyrmion solutions with cubic symmetry, whereas the
red-dashed dots correspond to the Skyrmion breaking up into
individual and weakly bound $B=1$ clumps, situated in a face-centered
cubic lattice (FCC) \cite{Gillard:2015eia}, see Appendix \ref{app:PT}. 

We clearly see that increasing the pion mass $m_\pi$, allows for
larger values of $\mathbf{m}_2^2$ and eventually for lower classical
binding energies.
We note, however, that $m_\pi=0.125$ has a smaller binding energy than
$m_\pi=0.25$ (corresponding to $m=1$ in
Refs.~\cite{Battye:2006na,Manko:2007pr}) before the symmetries change
from cubic to FCC.
Nevertheless, larger values of the pion mass parameter decrease the
binding energies; in particular, the classical binding energy is
smaller for $m_\pi=0.375$ and $m_\pi=0.5$ than for $m_\pi=0.125$. 
The largest value of $\mathbf{m}_2^2$ possible for cubic symmetry is
reached for $m_\pi=0.5$ and it also yields the smallest classical
binding energy of about 2.7\% for the 4-Skyrmion. 

We can also see from Fig.~\ref{fig:cbe} that the slope of the curves
at large $\mathbf{m}_2^2$ is much smaller than for small values; hence
the difference between the standard (linear) pion mass and the
modified pion mass becomes much less pronounced (recall that it
corresponds merely to a negative shift in the value of
$\mathbf{m}_2^2$).

\subsection{Calibration}

\begin{figure}[!htp]
  \begin{center}
    \includegraphics[width=0.75\linewidth]{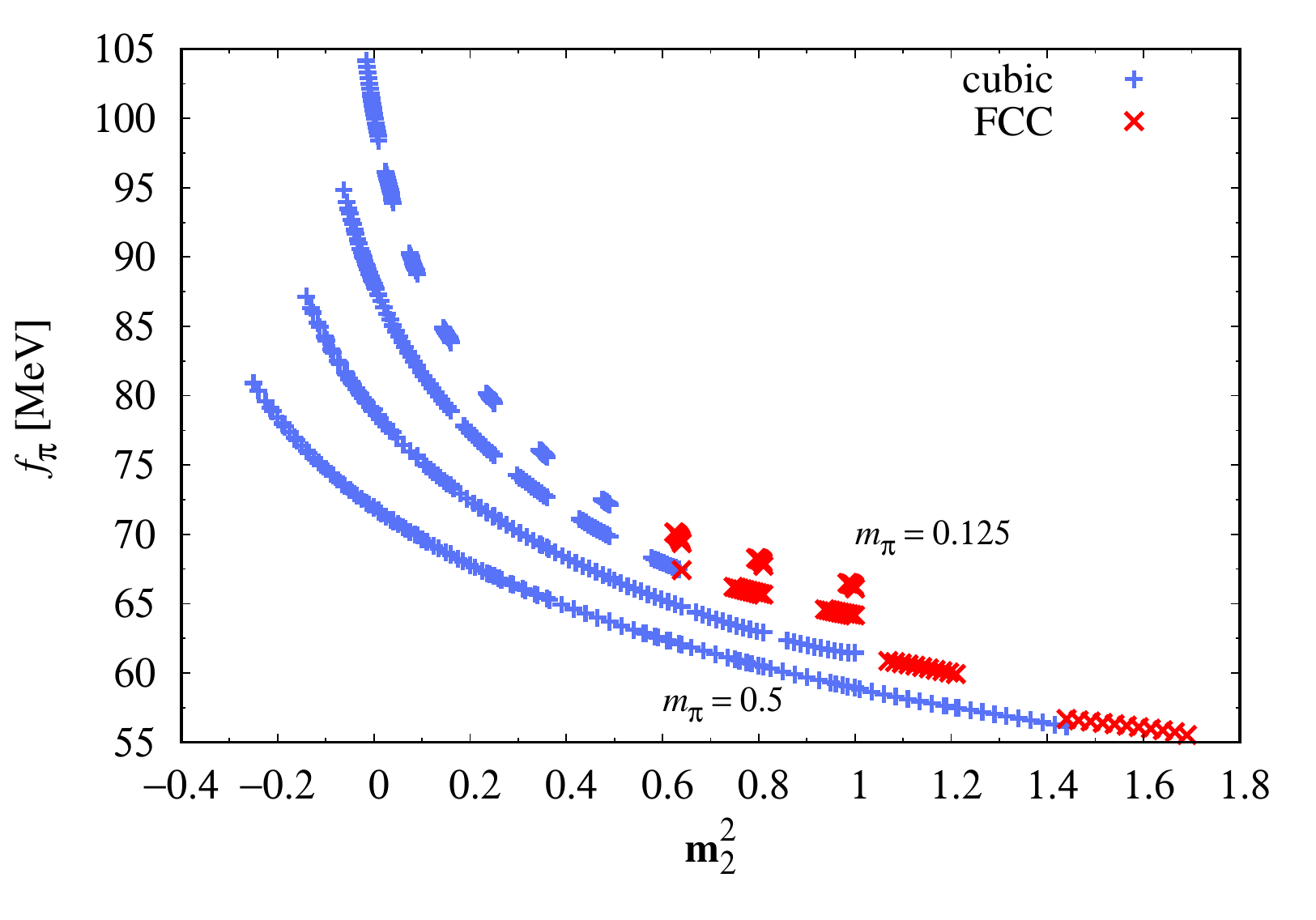}
    \caption{Calibration of the pion decay constant, $f_\pi$, as
      function of
      $\mathbf{m}_2^2$; four series of points are shown corresponding
      to $m_\pi=0.125,0.25,0.375,0.5$.}
    \label{fig:fpi}
  \end{center}
\end{figure}

\begin{figure}[!htp]
  \begin{center}
    \includegraphics[width=0.75\linewidth]{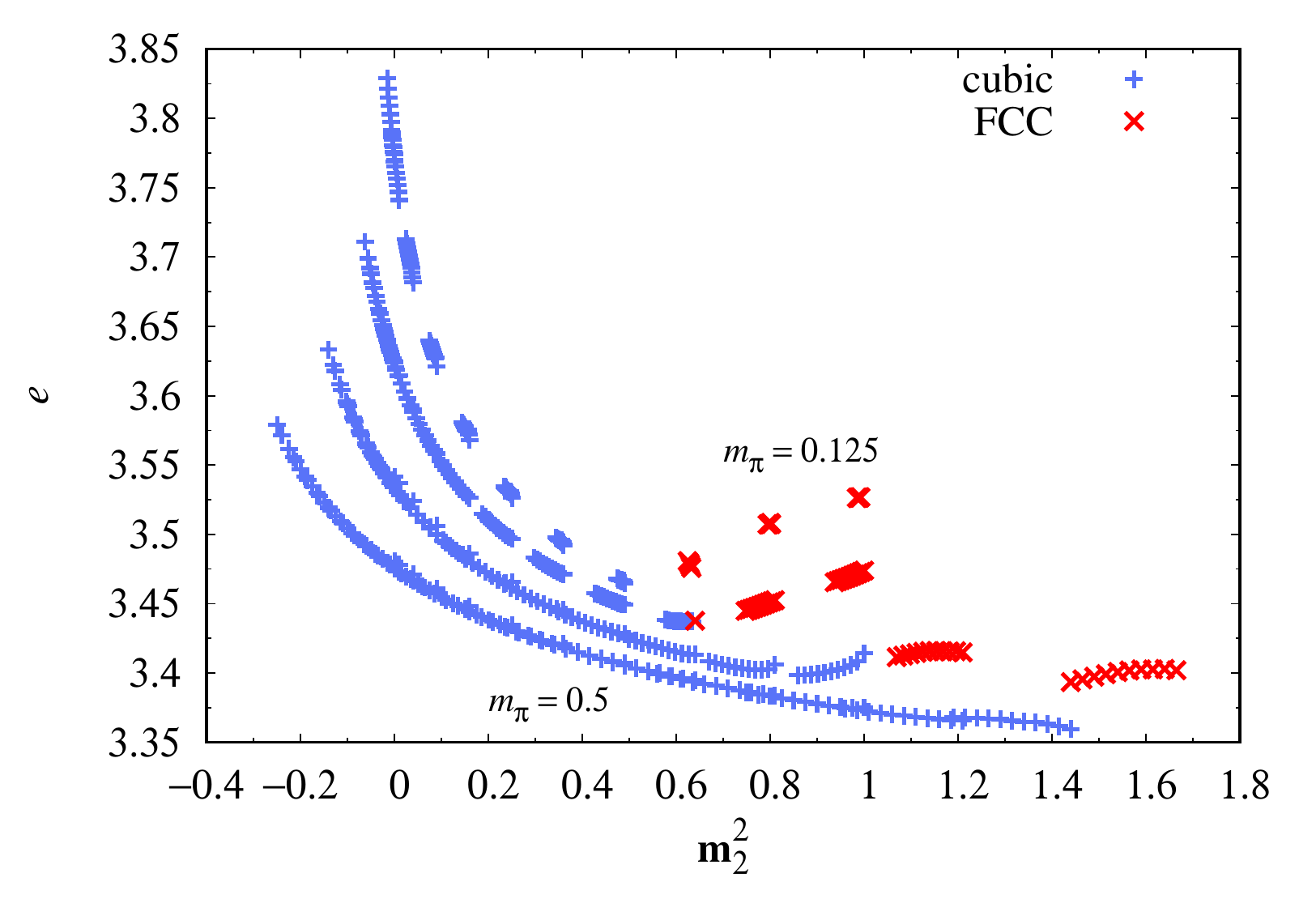}
    \caption{Calibration of the Skyrme term coefficient, $e$, as
      function of
      $\mathbf{m}_2^2$; four series of points are shown corresponding
      to $m_\pi=0.125,0.25,0.375,0.5$. }
    \label{fig:e}
  \end{center}
\end{figure}

Now we will perform a calibration to the ${}^4$He nucleus for each
Skyrmion solution in the parameter space.
In particular, as mentioned in Sec.~\ref{sec:model}, we fit the mass 
and the size of the 4-Skyrmion to those of ${}^4$He; this determines
$f_\pi$ and $e$. 
Figs.~\ref{fig:fpi} and \ref{fig:e} show the pion decay constant and
the (dimensionless) Skyrme term parameter $e$.

We can see from Fig.~\ref{fig:fpi} that for $m_2=0$, the modified pion
mass term -- corresponding to negative values of $\mathbf{m}_2^2$ --  
increases the pion decay constant (which is good).
The experimentally observed value is around 184 MeV (not shown in the
figure) in the normalization used in the Skyrme model
\cite{Adkins:1983ya}.
However, turning on $m_2$, which corresponds to positive values of
$\mathbf{m}_2^2$ (and decreases the binding energy) reduces the value
of $f_\pi$ to about a third of its experimentally measured value.
We observed that larger values of the pion mass directly translate
into smaller values of the pion decay constant, $f_\pi$. 

Since the value of the Skyrme term coefficient $e$ is to the best of
our knowledge not known experimentally, there is no preferred value;
it is simply the result of the fit of length and energy units.
Let us however remark that the blue points move downwards (decreasing
$e$) for increasing $\mathbf{m}_2^2$ (except for the largest values of 
the $m_\pi=0.375$ series before the symmetry changes from cubic to
FCC), whereas when the symmetry switches to FCC the red-dashed points
are moving upwards (increasing $e$) for increasing $\mathbf{m}_2^2$
(but possibly saturating at a plateau).
Let us also remark that the smaller $e$ is, the smaller the
contribution from spin-isospin quantization to the 1-Skyrmion is.
This means that in order to get the smallest possible total binding
energy, we need an as small as possible value of $e$.
For this, the large values of the pion mass $m_\pi$ are advantageous. 

Let us also remark that the curves of the binding energies shown in
Fig.~\ref{fig:cbe} are far smoother than those shown in
Fig.~\ref{fig:e}; this is due to the highly precise calculations for
the energies, whereas the sizes have been estimated without taking any
limits of large relaxation times.
This can be seen as small jumps in the curves of $e$ in
Fig.~\ref{fig:e}.
The error is however still smaller than or about the permille level.

\subsection{Mass spectrum}

\begin{figure}[!htp]
  \begin{center}
    \includegraphics[width=0.75\linewidth]{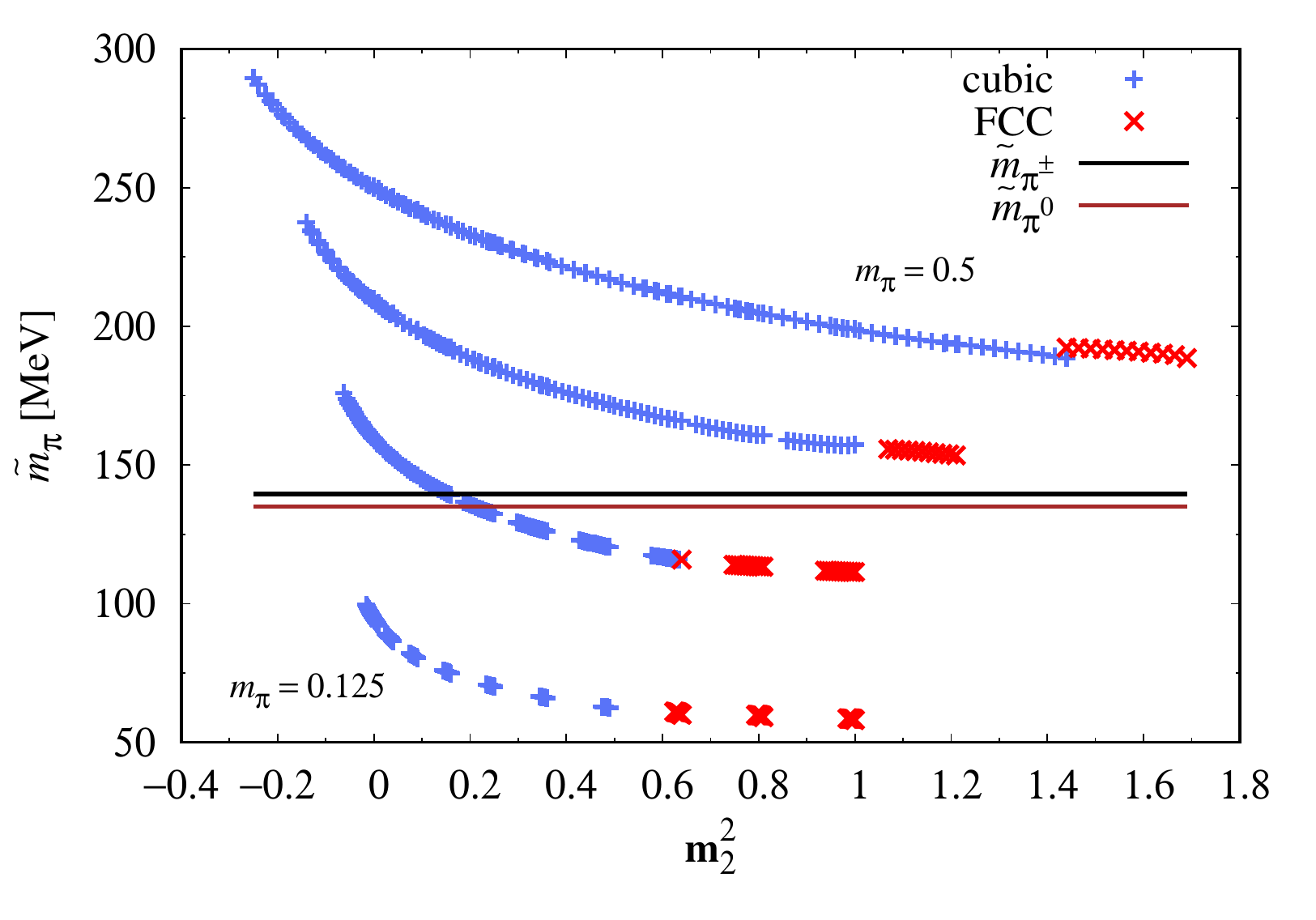}
    \caption{Pion mass in physical unit, $\tilde{m}_\pi$, as function
      of
      $\mathbf{m}_2^2$; four series of points are shown corresponding
      to $m_\pi=0.125,0.25,0.375,0.5$. }
    \label{fig:mpi}
  \end{center}
\end{figure}

\begin{figure}[!htp]
  \begin{center}
    \includegraphics[width=0.75\linewidth]{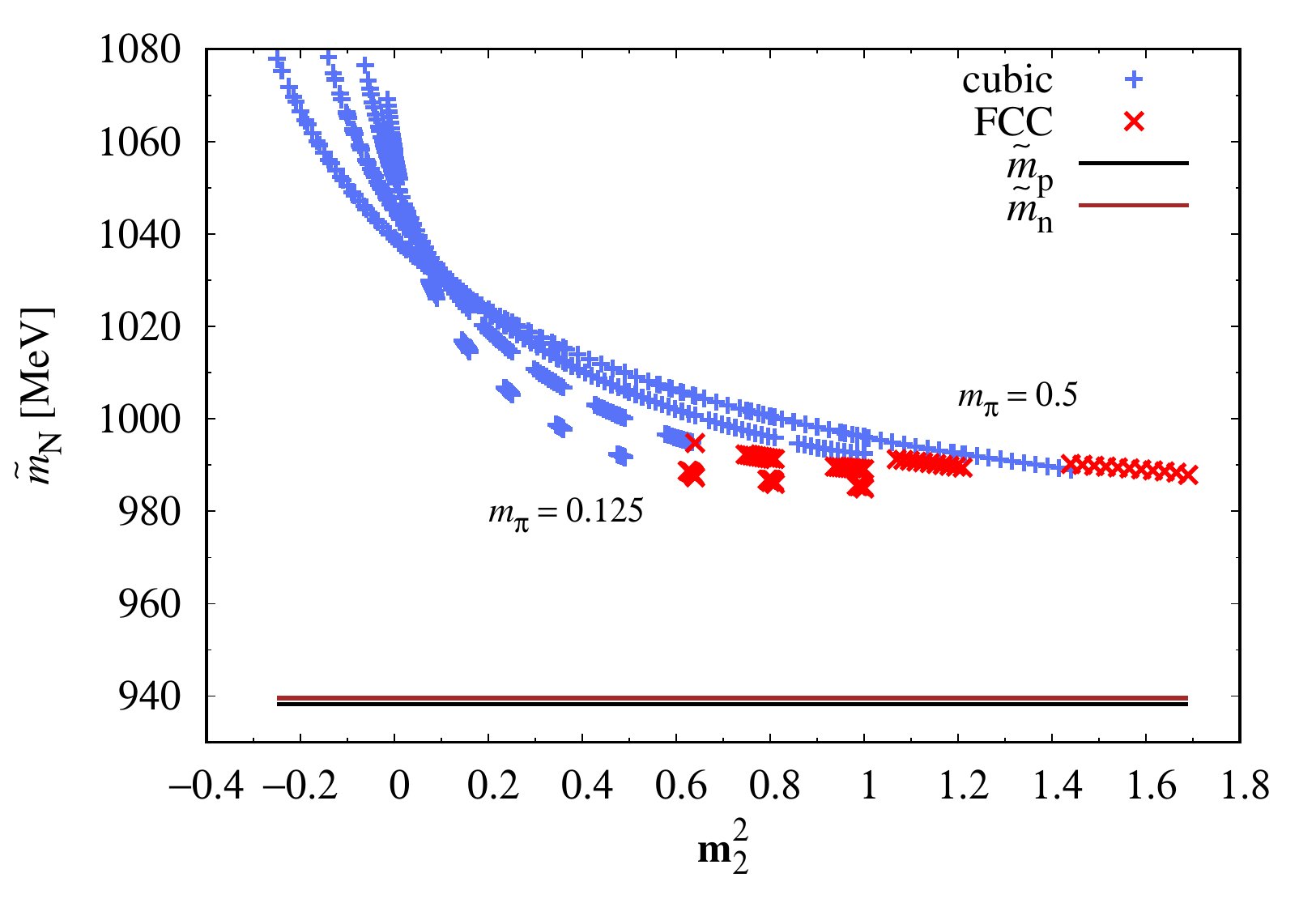}
    \caption{Nucleon mass in physical unit, $\tilde{m}_N$, as function
      of
      $\mathbf{m}_2^2$; four series of points are shown corresponding
      to $m_\pi=0.125,0.25,0.375,0.5$. }
    \label{fig:mn}
  \end{center}
\end{figure}

\begin{figure}[!htp]
  \begin{center}
    \includegraphics[width=0.75\linewidth]{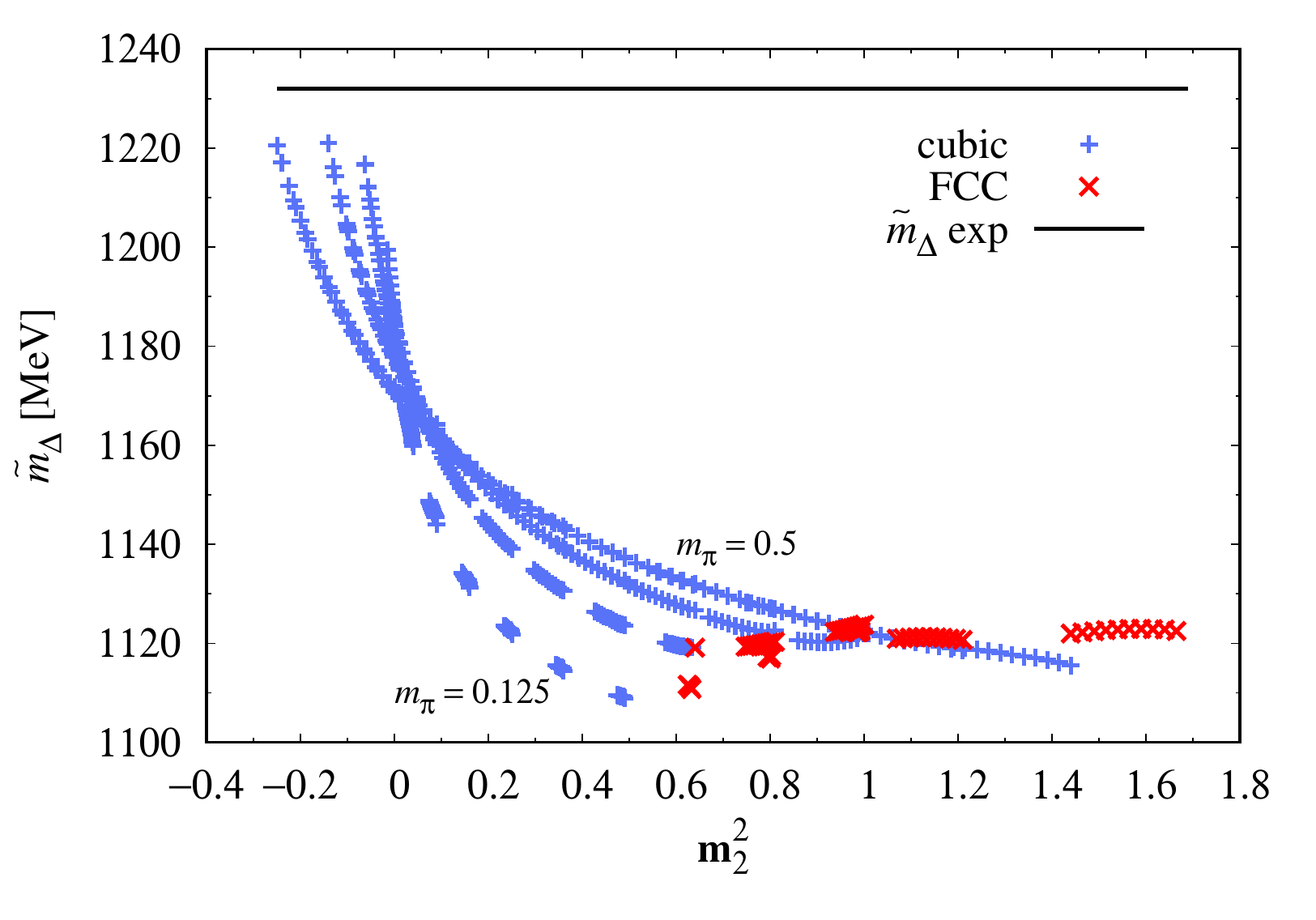}
    \caption{Mass of the $\Delta$ resonance in physical unit,
      $\tilde{m}_\Delta$, as function of
      $\mathbf{m}_2^2$; four series of points are shown corresponding
      to $m_\pi=0.125,0.25,0.375,0.5$. }
    \label{fig:mdelta}
  \end{center}
\end{figure}

We now turn to the mass spectrum.
Since the mass and size of the 4-Skyrmion has been fitted to that of
${}^4$He, $f_\pi$ and $e$ are fixed.
The masses in physical units of the pion, the nucleon and the delta
resonance can thus readily be calculated and they are presented in
Figs.~\ref{fig:mpi}, \ref{fig:mn} and \ref{fig:mdelta}, respectively.  

Let us start with the pion mass of Fig.~\ref{fig:mpi}.
We can see that if we want to minimize the classical relative binding
energy by maximizing $m_2$, then the experimentally preferred value of
$m_\pi$ is between $0.25$ and $0.375$; i.e.~between the second and the
third series in Fig.~\ref{fig:mpi}. 
However, as already mentioned, since the pion decay constant is almost
a factor of 3 off of its experimental value and the fact that we
choose to interpret $f_\pi$ and $m_\pi$ as renormalized constants in
the baryon medium, not in the pion vacuum, then we may contemplate
allowing for some slush also in the value of $m_\pi$.

Looking now at Fig.~\ref{fig:mn}, interestingly, we can see that 
for the largest possible value of $\mathbf{m}_2$, the nucleon mass is
closest to the experimentally observed value for all of the pion mass
values.
The pion mass $m_\pi=0.5$ gives slightly better, but nearly the same
value as for $m_\pi=0.125$ and for $m_\pi=0.25$ the worst fit to the
measured nucleon mass is found, in the limit of the largest possible
value of $\mathbf{m}_2^2$ before the cubic symmetry is lost.
For all points in the parameter space, we can conclude that the
loosely bound Skyrme model overestimates the nucleon mass. 

The final mass we calculate in this paper is the mass of the delta 
resonance.
We can see from Fig.~\ref{fig:mdelta} that for the largest
possible values of $\mathbf{m}_2^2$, for each series, the model
estimate is the farthest away from the measured mass of the delta
resonance.  
For all points in the parameter space, we can conclude that the
loosely bound Skyrme model underestimates the mass of the delta
resonance.

\subsection{Proton charge radius}

\begin{figure}[!htp]
  \begin{center}
    \includegraphics[width=0.75\linewidth]{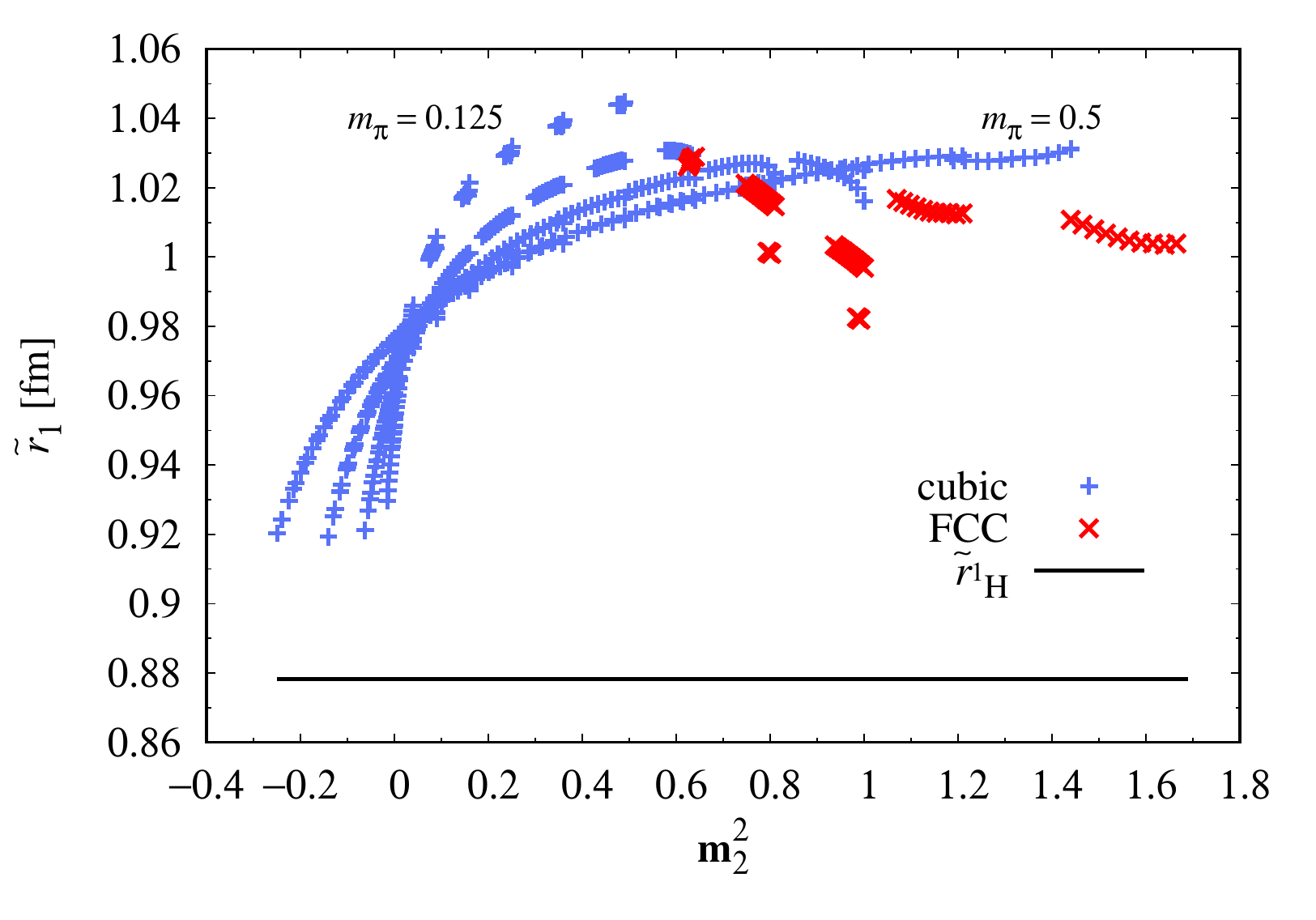}
    \caption{Charge radius of the proton in physical unit,
      $\tilde{r}_1$, as function of
      $\mathbf{m}_2^2$; four series of points are shown corresponding
      to $m_\pi=0.125,0.25,0.375,0.5$. }
    \label{fig:rn}
  \end{center}
\end{figure}

We will now turn to the proton charge radius and use it as a rough
estimate of the size of the nucleon.
Fig.~\ref{fig:rn} shows the square root of the squared radius averaged 
using the baryon charge density\footnote{Ref.~\cite{Adam:2015zhc}
  argues that the baryon charge density is a natural definition for
  calculating the size of a soliton; in particular in Skyrme-type
  models.  } of the 1-Skyrmion, see Eq.~\eqref{eq:rsqBdensity}. 

We can observe from the figure that all the proton charge radii in the
entire parameter space are overestimated. This is because we fit the
length scale to the size of ${}^4$He and the 4-Skyrmion in general is
too small; the addition of the loosely bound potential, $m_2>0$, in
turn exacerbates this tendency and decreases the 4-Skyrmion even
more. 
With this choice of fitting, this problem shows up as the charge
radius for the proton being too large.
We can, interestingly, observe that if the loosely bound potential is
turned off ($m_2=0$), then the modified pion mass improves the value of
the charge radius (corresponding to negative values of
$\mathbf{m}_2^2$). 
This is, however, in the part of the parameter space where the
classical relative binding energies for the 4-Skyrmion are the largest
and hence most at odds with experimental data.

Another effect that we can observe from Fig.~\ref{fig:rn} is that
before the threshold for cubic symmetry is reached, i.e.~the boundary
between when the symmetry of the 4-Skyrmion is cubic or FCC, then the
charge radius increases (except for $m_\pi=0.375$).
However, once the symmetry has changed to FCC, the size of the
4-Skyrmion increases quite a lot and this in turn has the effect of
reducing the charge radius of the proton (because we fit the length
scale to the size of the 4-Skyrmion); in fact, for increasing
$\mathbf{m}_2^2$, the red-dashed points move downwards in the figure.

\subsection{Total binding energies}

\begin{figure}[!htp]
  \begin{center}
    \includegraphics[width=0.75\linewidth]{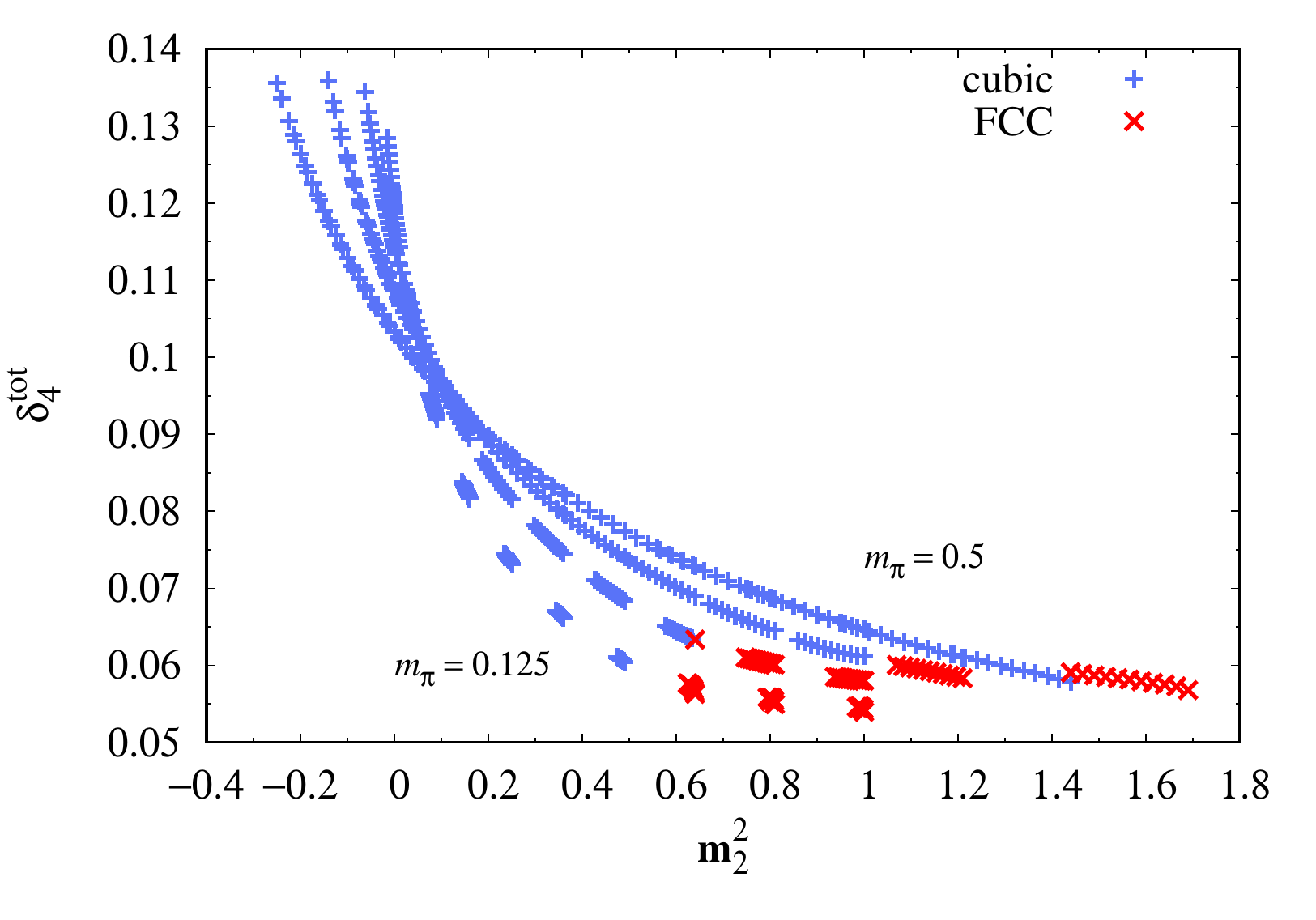}
    \caption{Total binding energy, $\delta_4^{\rm tot}$, as
      function of
      $\mathbf{m}_2^2$; four series of points are shown corresponding
      to $m_\pi=0.125,0.25,0.375,0.5$. }
    \label{fig:tbe}
  \end{center}
\end{figure}

\begin{figure}[!htp]
  \begin{center}
    \mbox{
      \subfloat[$m_\pi=0.125$]{\includegraphics[width=0.49\linewidth]{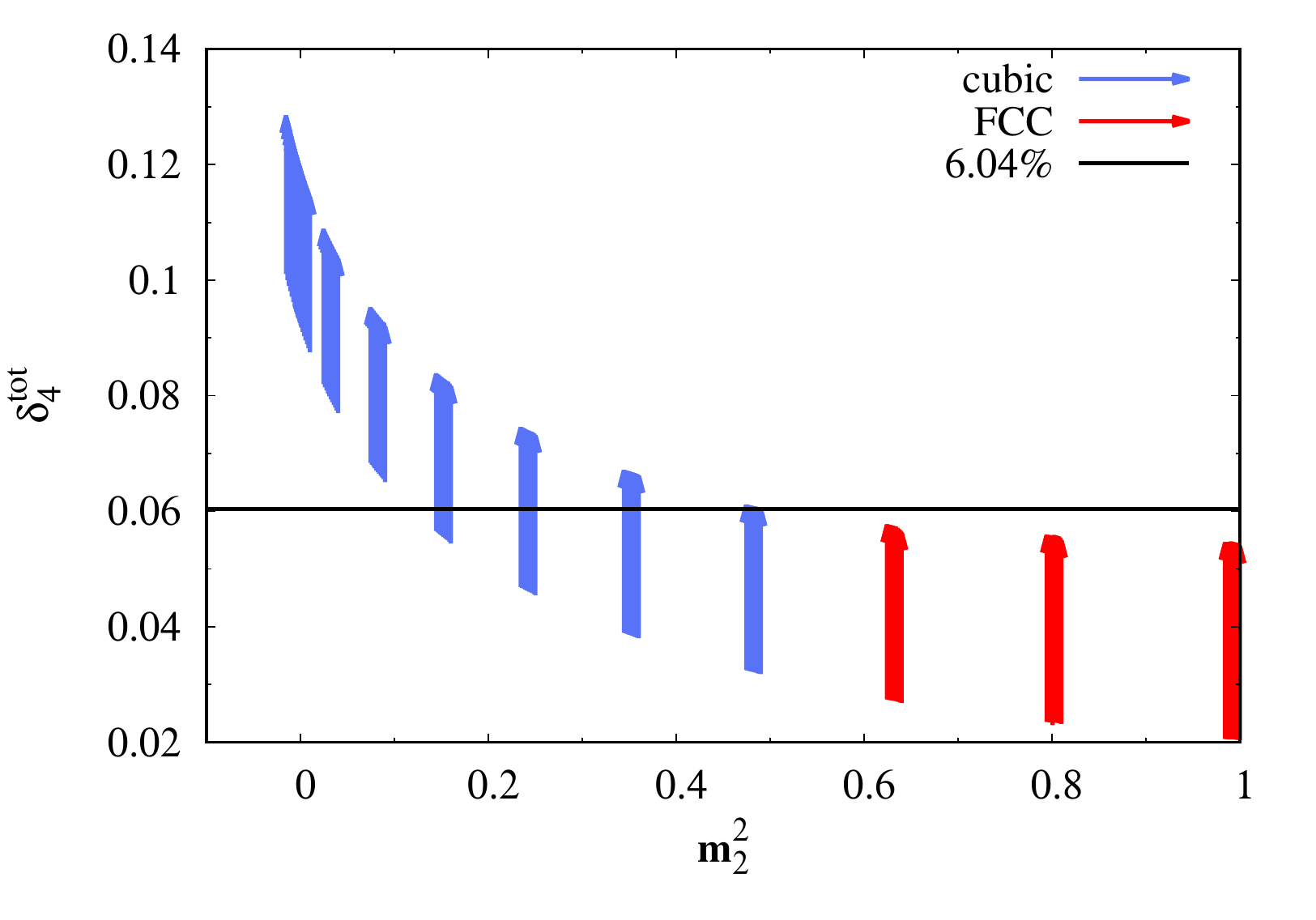}}
      \subfloat[$m_\pi=0.25$]{\includegraphics[width=0.49\linewidth]{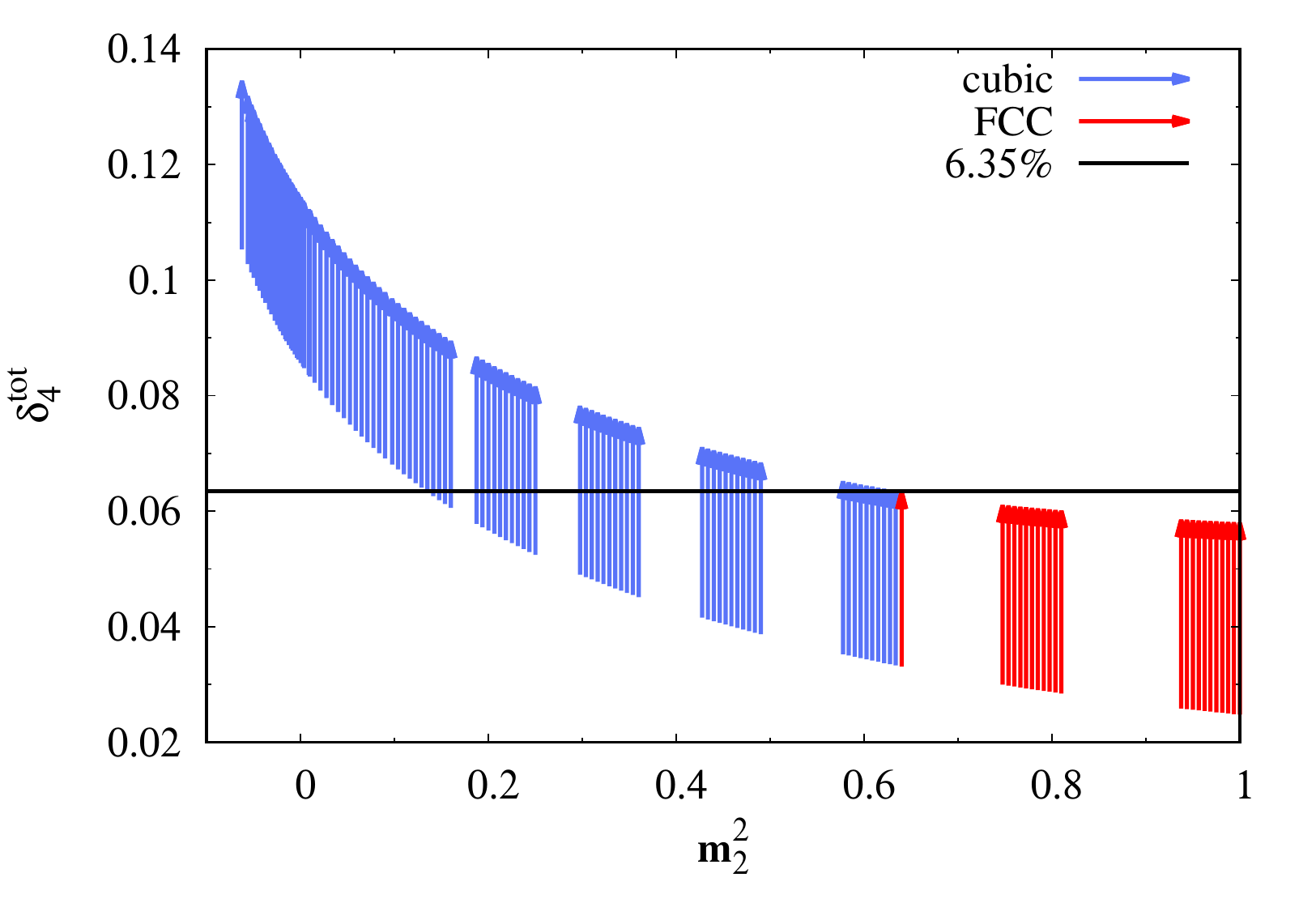}}}
    \mbox{
      \subfloat[$m_\pi=0.375$]{\includegraphics[width=0.49\linewidth]{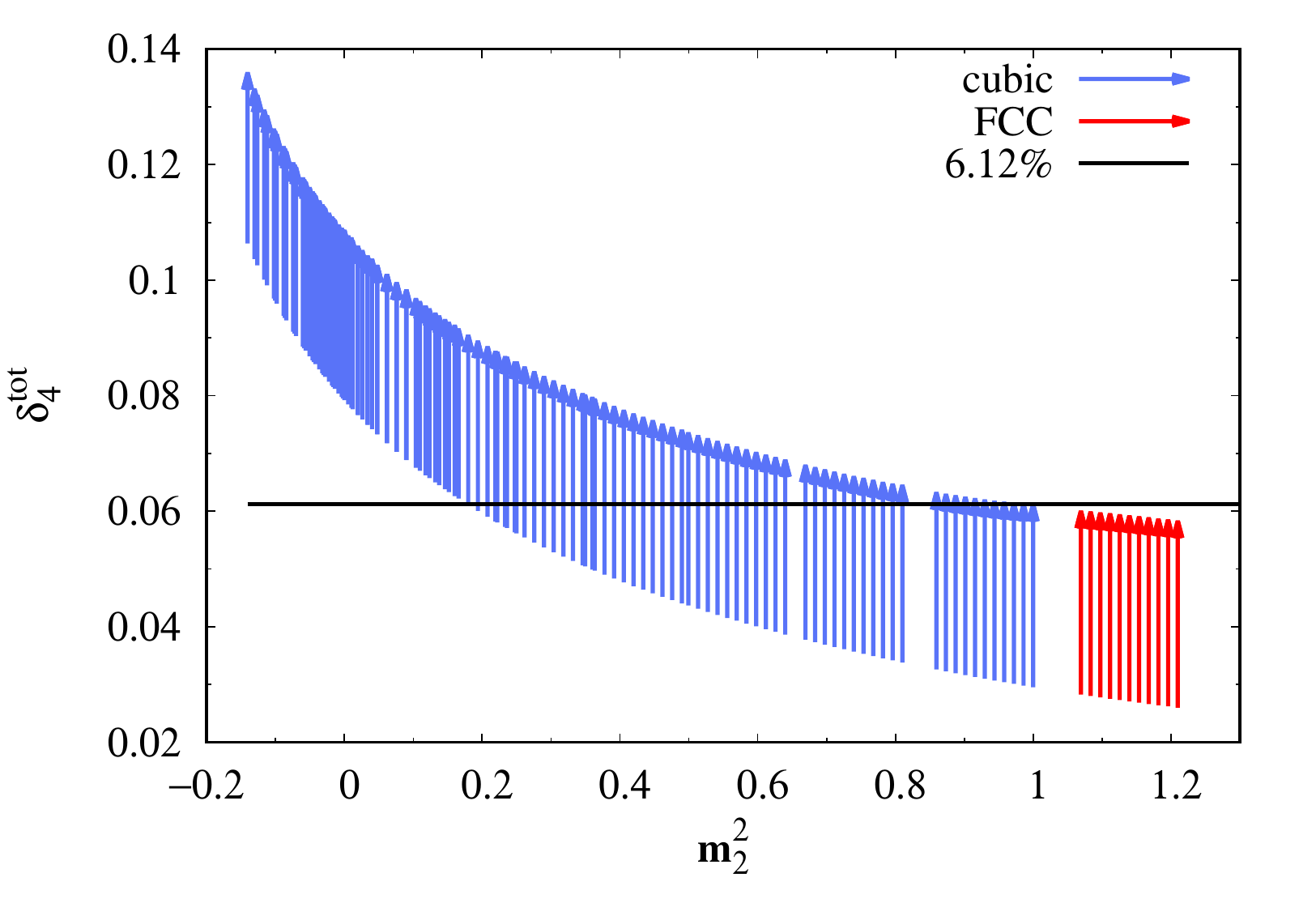}}
      \subfloat[$m_\pi=0.5$]{\includegraphics[width=0.49\linewidth]{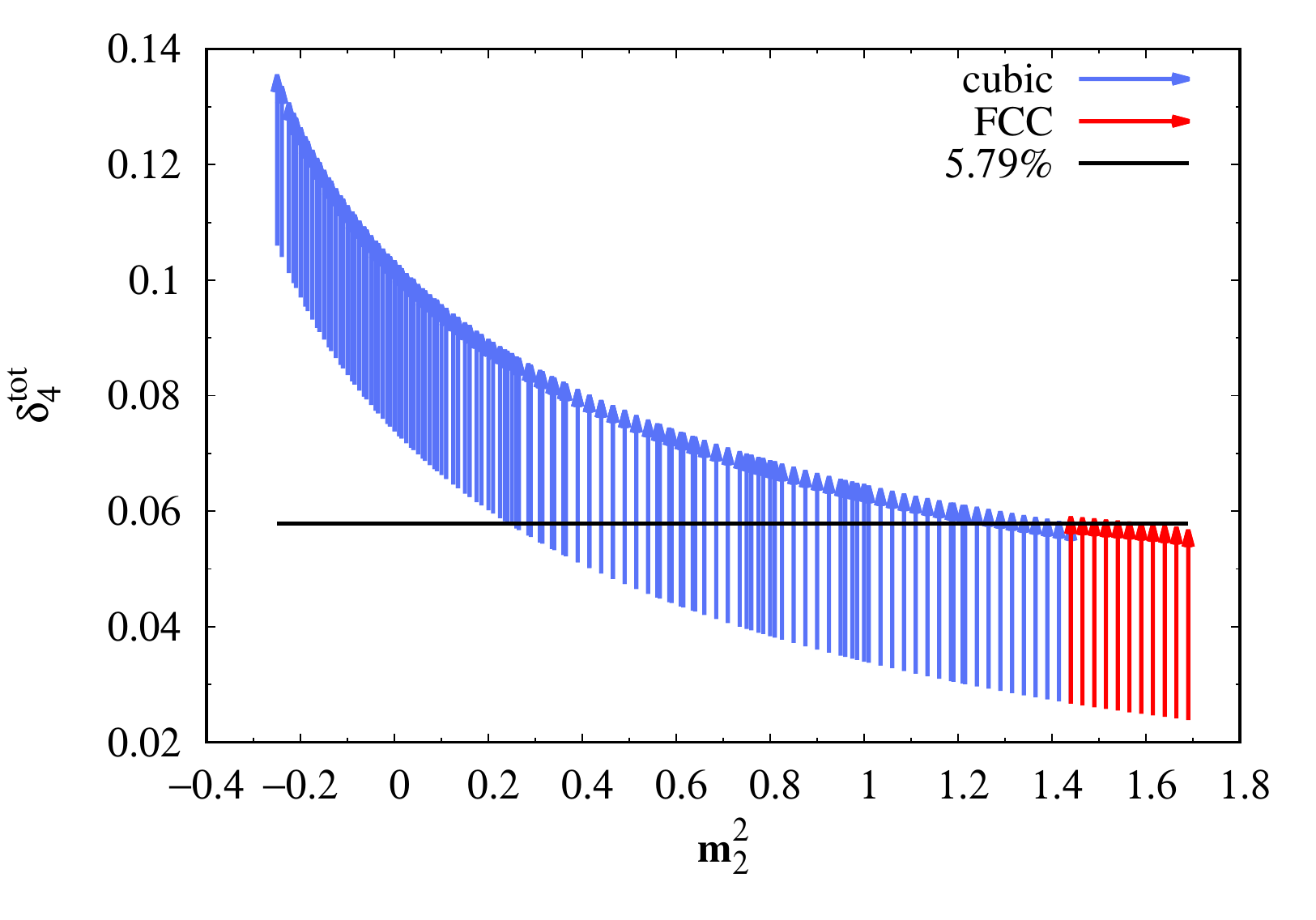}}}
    \caption{Breakdown of the total binding energy, $\delta_4^{\rm tot}$, as
      function of
      $\mathbf{m}_2^2$; four series of points are shown corresponding
      to $m_\pi=0.125,0.25,0.375,0.5$.
      The bottom of the arrows corresponds to the classical contribution;
      whereas the arrow head includes the quantum correction. }
    \label{fig:tbebd}
  \end{center}
\end{figure}

The final comparison with experiment is again the relative binding
energy, but now we will take the quantum contribution due to
spin-isospin quantization into account.
The total relative binding energy is defined in
Eq.~\eqref{eq:delta4tot}.
Fig.~\ref{fig:tbe} shows the total relative binding energy and
Fig.~\ref{fig:tbebd} displays the breakdown of the classical
contribution (the bottom of the arrows) and the quantum contribution 
(the length of the arrows). 

As was the case for the classical relative binding energy, so is the
case for the total relative binding energy; the loosely bound
potential decreases the binding energy.
We can see that the lowest binding energy is reached for $m_\pi=0.5$,
but the next-to-best value is for $m_\pi=0.125$; the dependence of
the total binding energy on the pion mass parameter is not linear and
indeed quite nontrivial.

From Fig.~\ref{fig:tbebd} we can see that the quantum contribution
increases slightly when the loosely bound potential is turned on;
i.e.~when $\mathbf{m}_2^2$ is large.

\section{Discussion and conclusion}\label{sec:discussion}

In this paper we have found that the modified pion mass increases the
binding energy of the $B$-Skyrmions as one would expect.
We also found that the cubic symmetry is kept for slightly larger
values of the coefficient of the loosely bound potential when the
modified pion mass term is used, compared to when the standard pion
mass term is used.
This is because a given value of the modified pion mass term
corresponds to the same standard pion mass albeit with a reduce value
of the loosely bound mass parameter $\mathbf{m}_2^2=m_2^2-m_\pi^2$. 
We found -- as pointed out many places in the literature -- that the
model prefers quite large values of the pion mass; this allows us to
use a larger coefficient of the loosely bound potential and hence
reduce the binding energy further.
We are able to reduce the classical binding energy to about the 2.7\%
level and the total binding energy to about the 5.8\% level.
This corresponds, however, to a rather large pion mass at 190 MeV, a
rather small pion decay constant at 56 MeV, a nucleon mass at 990 MeV,
the mass of the delta resonance at 1118 MeV and finally a charge
radius of the proton at 0.97 fm. 

This systematic study has only lowered the relative binding energy
by about 0.6\% with respect to that found in
Ref.~\cite{Gudnason:2016mms}.
However, in this spirit of systematically surveying the parameter
space of the Skyrme model, there are plenty of directions to look for
improvements.
One next step is to consider the BPS-Skyrme term; however, as we
mentioned in the introduction, its introduction to the model with a
large coefficient has proven notoriously difficult at the technical
level of numerical calculations.
Naturally one can extend this systematic study to the complete
potential of third order in $\sigma$.
Other effects that we would like to include in the future is the
breaking of the isospin symmetry and the Coulomb potential -- which
should be most significant for larger nuclei.

\subsection*{Acknowledgments}

S.~B.~G.~thanks the Recruitment Program of High-end Foreign
Experts for support.
The work of M.~N.~is supported in part by a Grant-in-Aid for
Scientific Research on Innovative Areas ``Topological Materials
Science'' (KAKENHI Grant No.~15H05855) and ``Nuclear Matter in Neutron
Stars Investigated by Experiments and Astronomical Observations''
(KAKENHI Grant No.~15H00841) from the the Ministry of Education,
Culture, Sports, Science (MEXT) of Japan. The work of M.~N.~is also
supported in part by the Japan Society for the Promotion of Science
(JSPS) Grant-in-Aid for Scientific Research (KAKENHI Grant
No.~16H03984) and by the MEXT-Supported Program for the Strategic
Research Foundation at Private Universities ``Topological Science''
(Grant No.~S1511006).

\appendix

\section{The cubic to FCC transition}\label{app:PT}

In this Appendix we show figures of the Skyrmions around the region in
parameter space where the Skyrmion changes symmetry from cubic
(platonic) symmetry to FCC symmetry.
The Skyrmion thus changes from being composed of eight half-Skyrmions
situated at the corners of a cube to being composed by four spheres 
sitting on the vertices of a tetrahedron. 

\begin{figure}[!thp]
  \centering
  \begin{minipage}[t]{0.47\linewidth}
    \mbox{
      \includegraphics[height=0.0795\textheight]{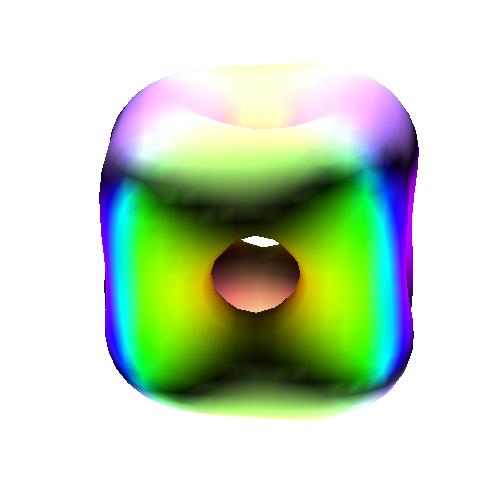}
      \includegraphics[height=0.0795\textheight]{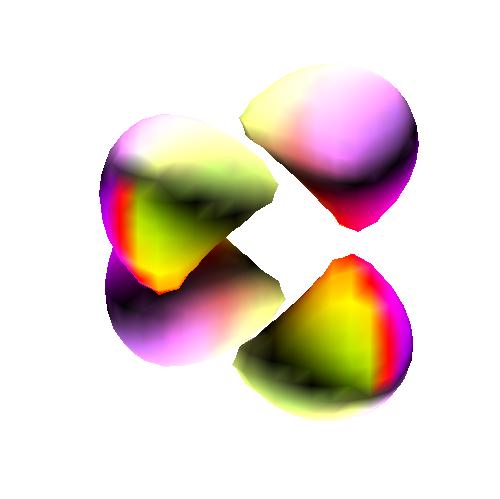}
      \includegraphics[height=0.0795\textheight]{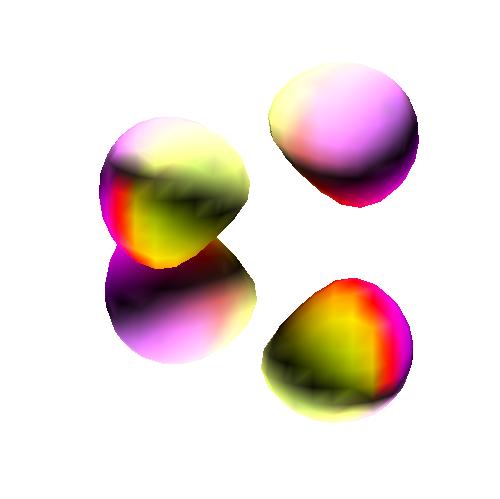}}\\
    \mbox{
      \includegraphics[height=0.0795\textheight]{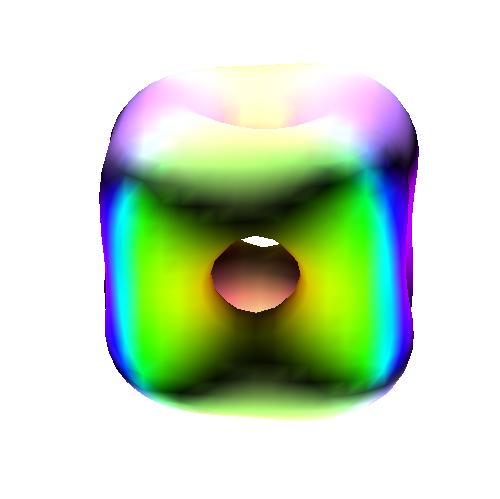}
      \includegraphics[height=0.0795\textheight]{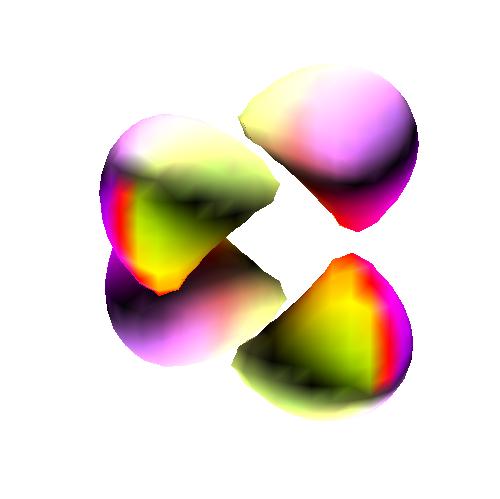}
      \includegraphics[height=0.0795\textheight]{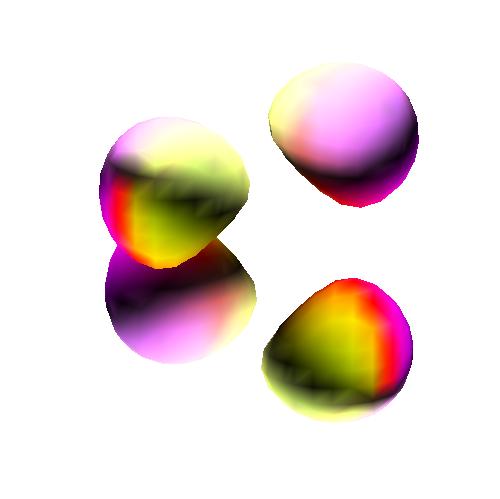}}\\
    \mbox{
      \includegraphics[height=0.0795\textheight]{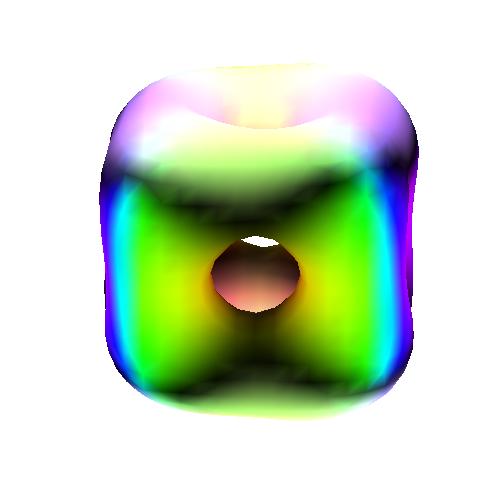}
      \includegraphics[height=0.0795\textheight]{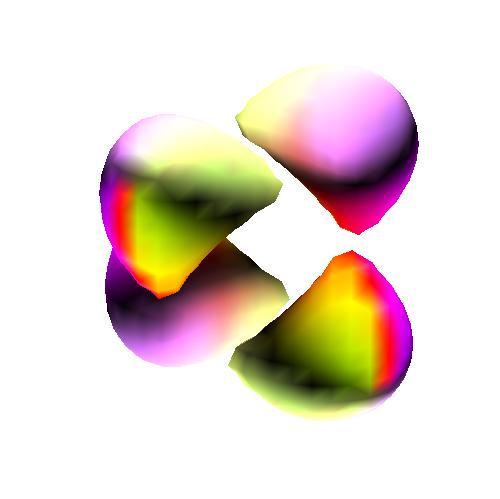}
      \includegraphics[height=0.0795\textheight]{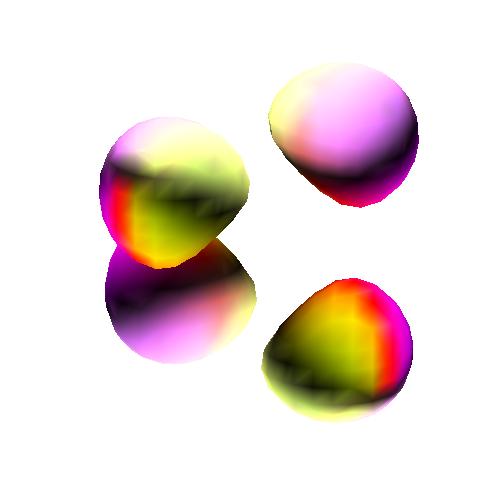}}\\
    \mbox{
      \includegraphics[height=0.0795\textheight]{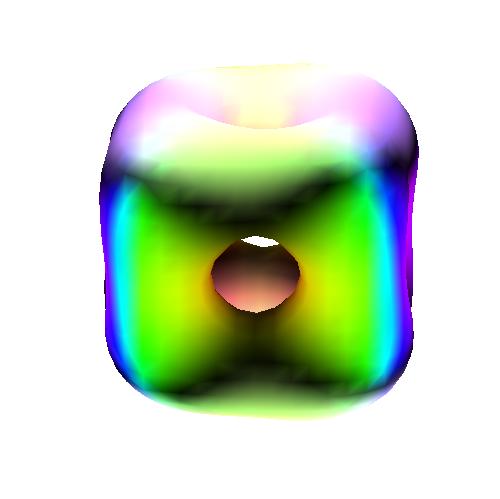}
      \includegraphics[height=0.0795\textheight]{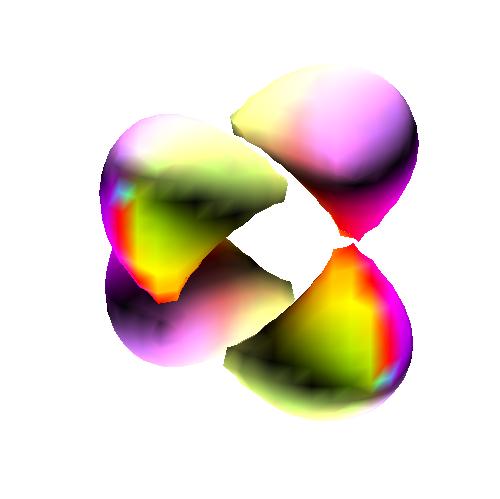}
      \includegraphics[height=0.0795\textheight]{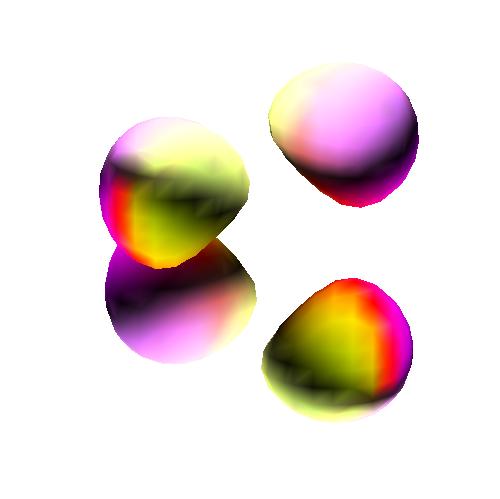}}\\
    \mbox{
      \includegraphics[height=0.0795\textheight]{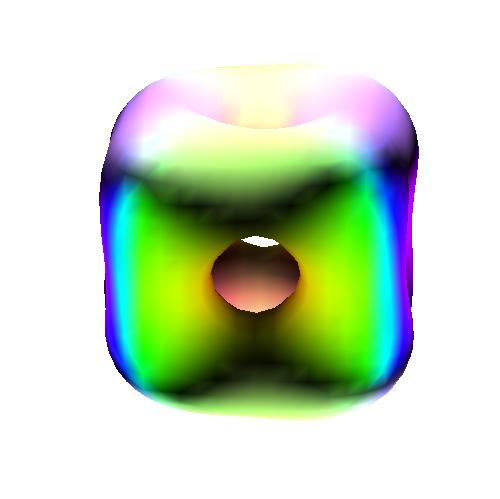}
      \includegraphics[height=0.0795\textheight]{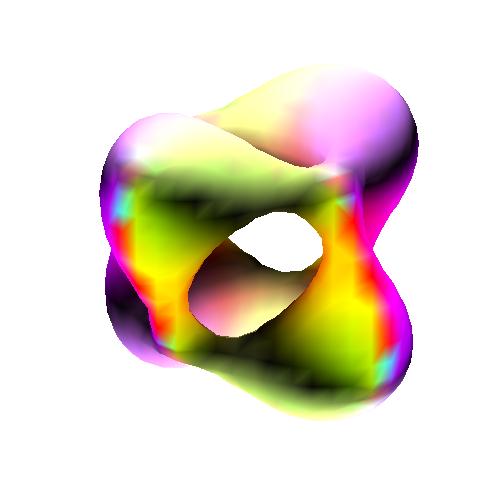}
      \includegraphics[height=0.0795\textheight]{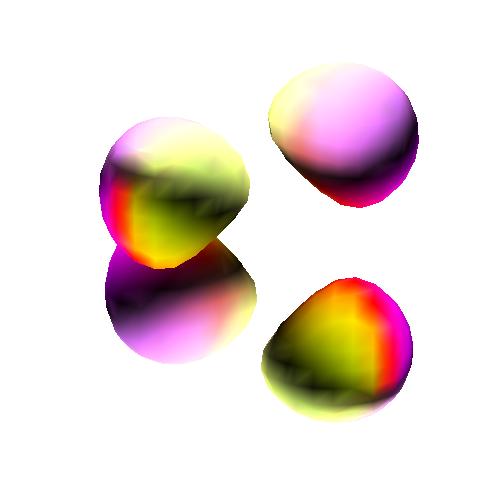}}\\
    \mbox{
      \includegraphics[height=0.0795\textheight]{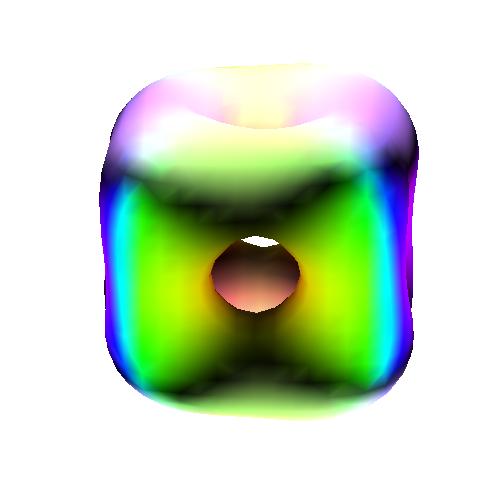}
      \includegraphics[height=0.0795\textheight]{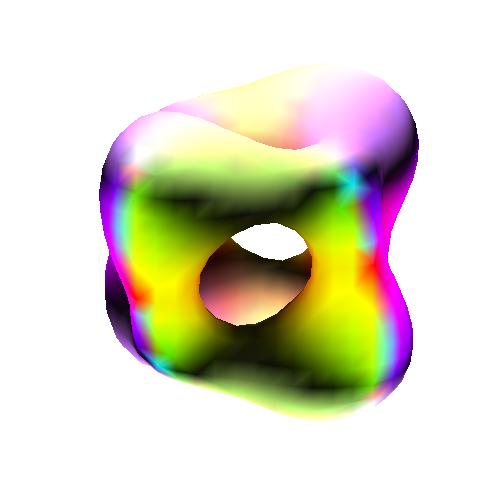}
      \includegraphics[height=0.0795\textheight]{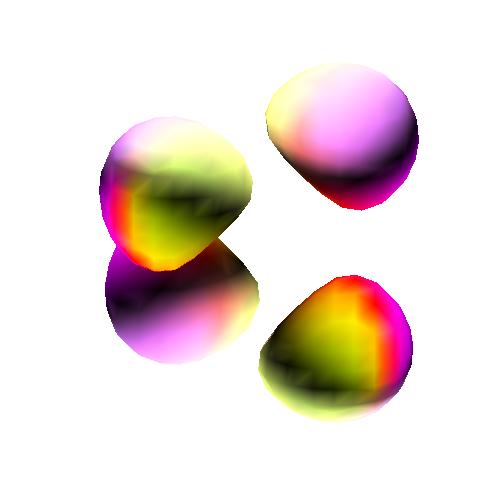}}
    \caption{The columns show $m_\pi=0.125$ Skyrmion solutions with
      $m_2=0.7,0.8,0.9$ and the rows correspond to $\alpha$ from 0 to
      1 in steps of 0.2 from top to bottom.}
    \label{fig:nout0125}
  \end{minipage}\ \ \
  \begin{minipage}[t]{0.47\linewidth}
    \mbox{
      \includegraphics[height=0.0795\textheight]{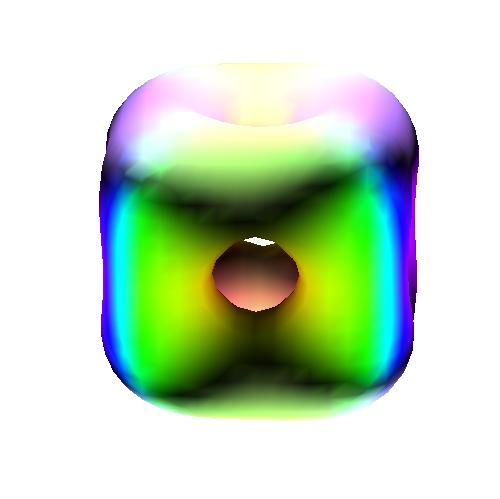}
      \includegraphics[height=0.0795\textheight]{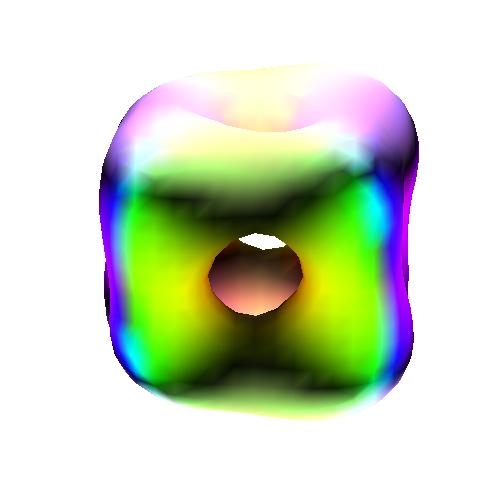}
      \includegraphics[height=0.0795\textheight]{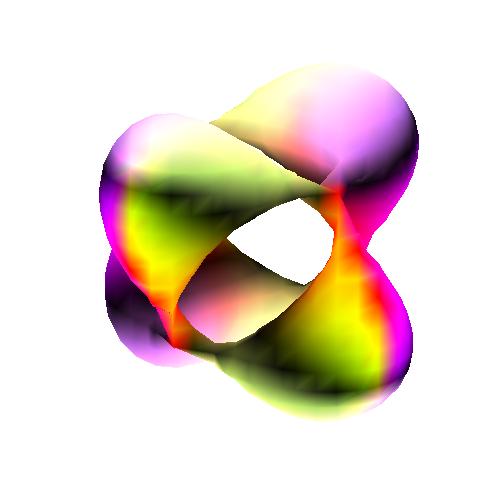}}\\
    \mbox{
      \includegraphics[height=0.0795\textheight]{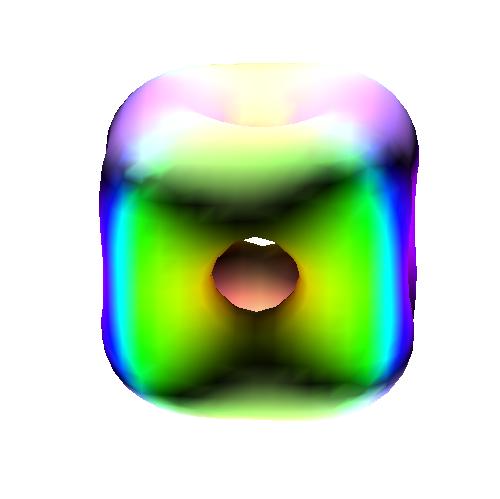}
      \includegraphics[height=0.0795\textheight]{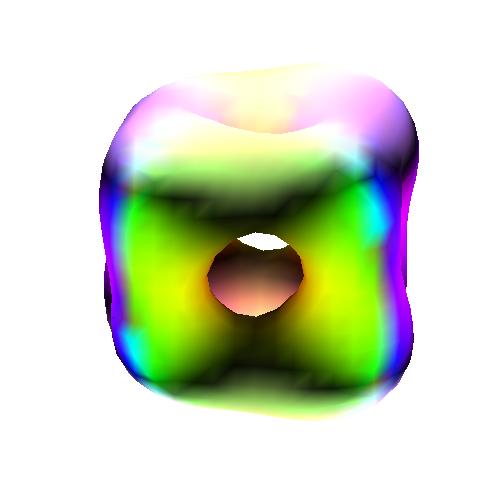}
      \includegraphics[height=0.0795\textheight]{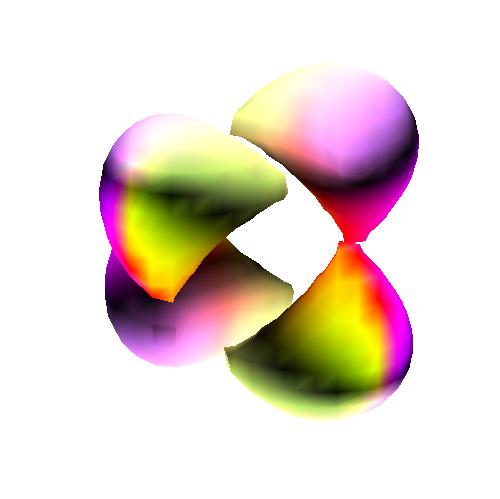}}\\
    \mbox{
      \includegraphics[height=0.0795\textheight]{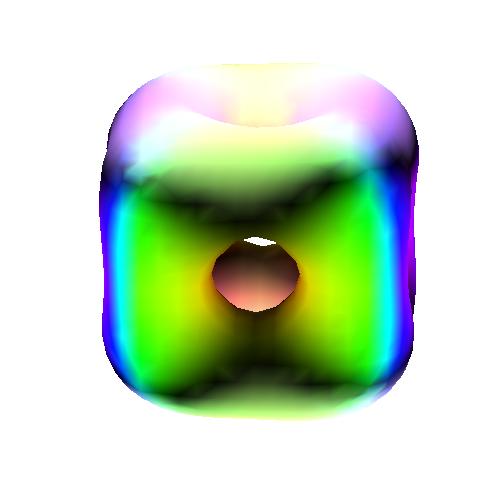}
      \includegraphics[height=0.0795\textheight]{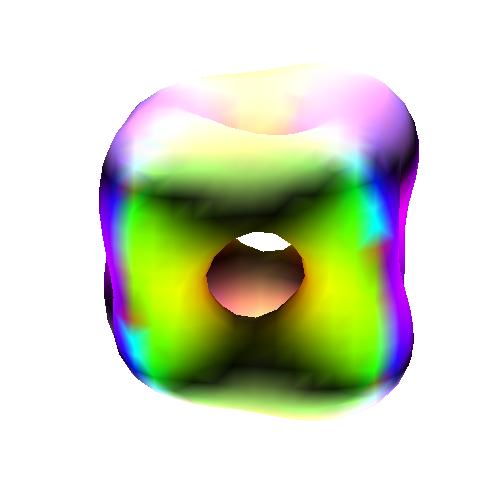}
      \includegraphics[height=0.0795\textheight]{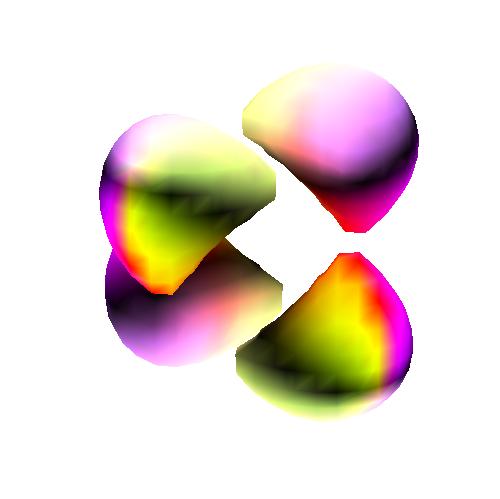}}\\
    \mbox{
      \includegraphics[height=0.0795\textheight]{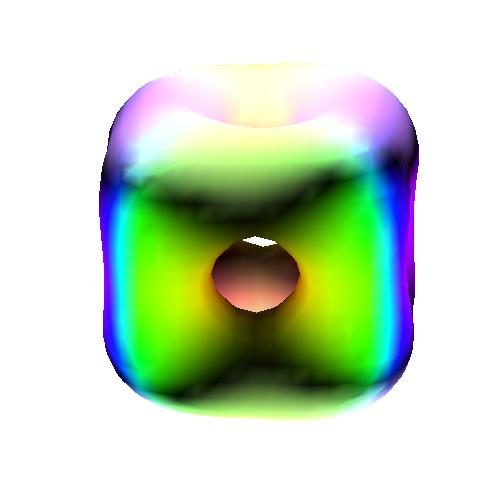}
      \includegraphics[height=0.0795\textheight]{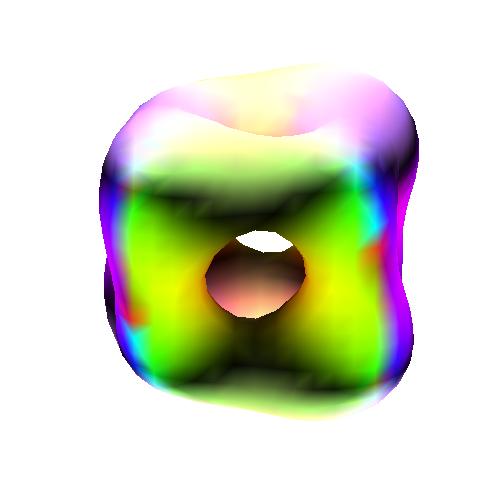}
      \includegraphics[height=0.0795\textheight]{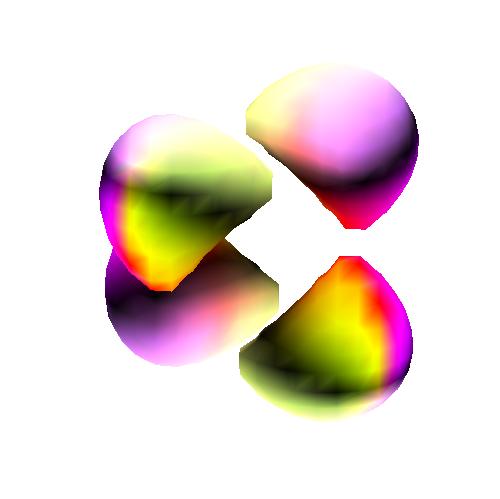}}\\
    \mbox{
      \includegraphics[height=0.0795\textheight]{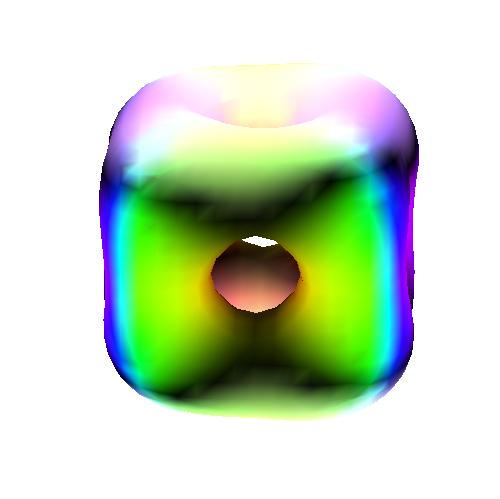}
      \includegraphics[height=0.0795\textheight]{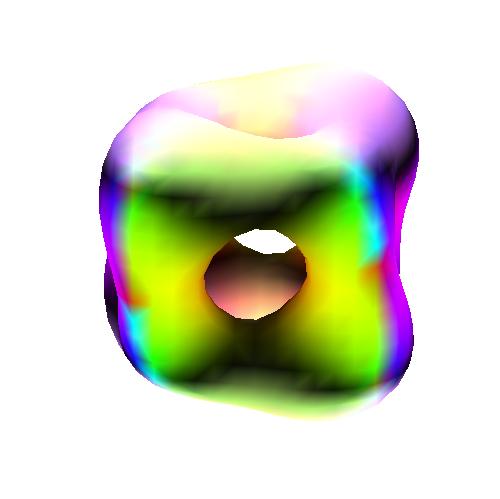}
      \includegraphics[height=0.0795\textheight]{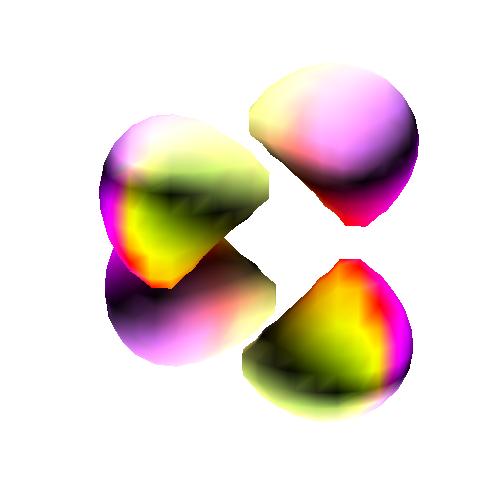}}\\
    \mbox{
      \includegraphics[height=0.0795\textheight]{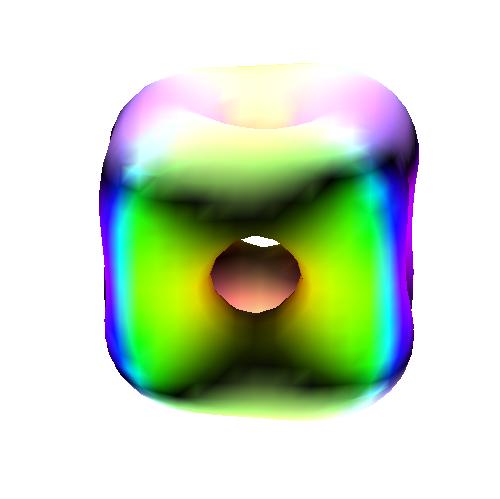}
      \includegraphics[height=0.0795\textheight]{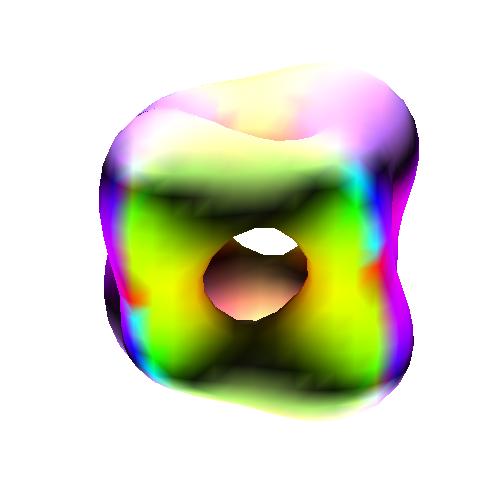}
      \includegraphics[height=0.0795\textheight]{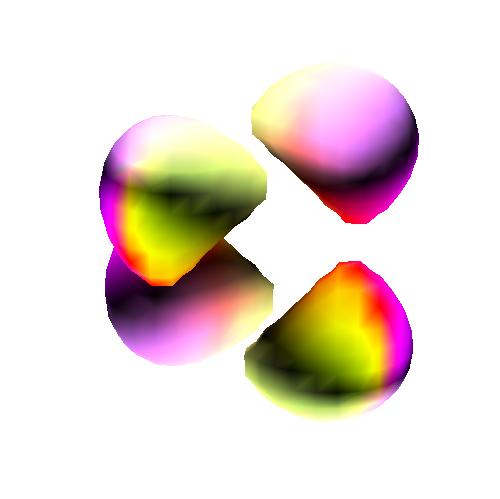}}
    \caption{The columns show $m_\pi=0.25$ Skyrmion solutions with
      $m_2=0.7,0.8,0.9$ and the rows correspond to $\alpha$ from 0 to
      1 in steps of 0.2 from top to bottom.
    }
    \label{fig:nout025}
\end{minipage}
\end{figure}

\begin{figure}[!p]
  \centering
  \begin{minipage}[t]{0.47\linewidth}
    \mbox{
      \includegraphics[height=0.0795\textheight]{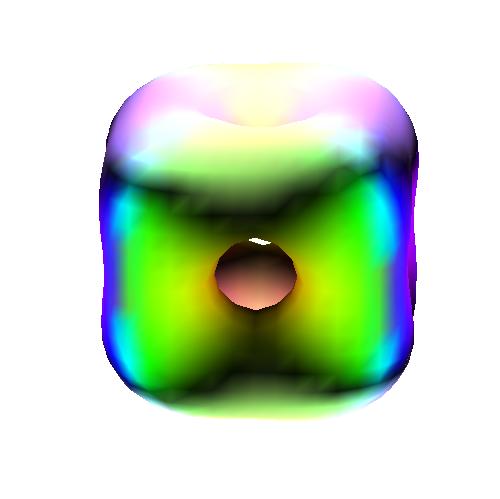}
      \includegraphics[height=0.0795\textheight]{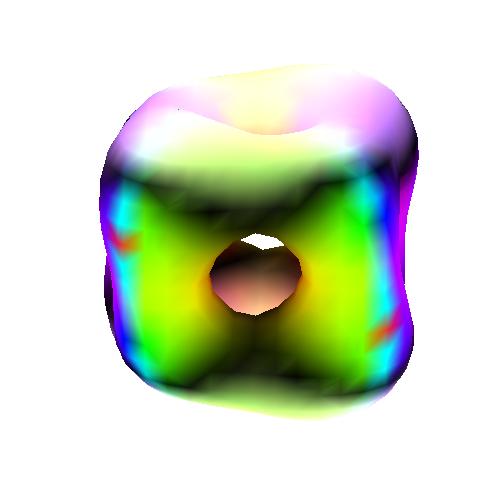}
      \includegraphics[height=0.0795\textheight]{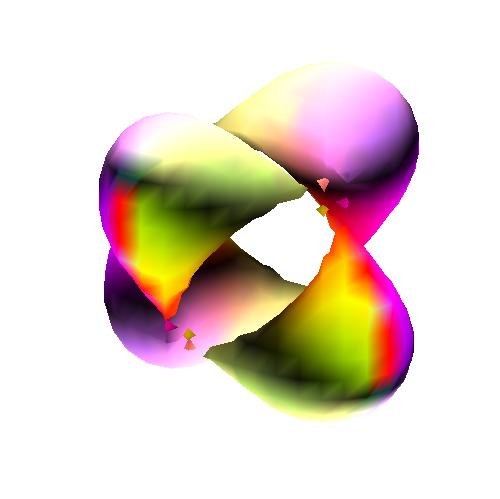}}\\
    \mbox{
      \includegraphics[height=0.0795\textheight]{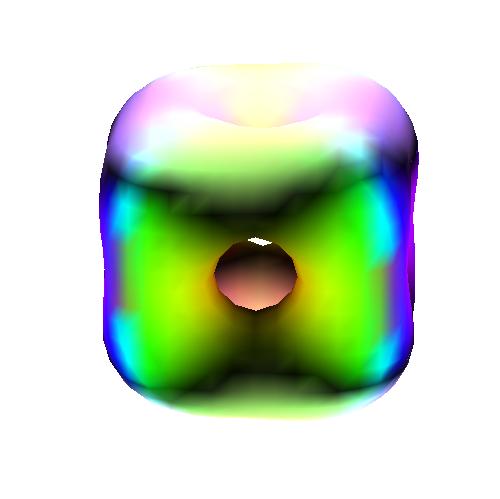}
      \includegraphics[height=0.0795\textheight]{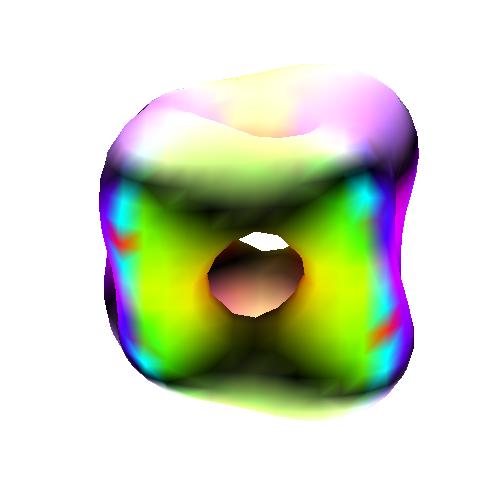}
      \includegraphics[height=0.0795\textheight]{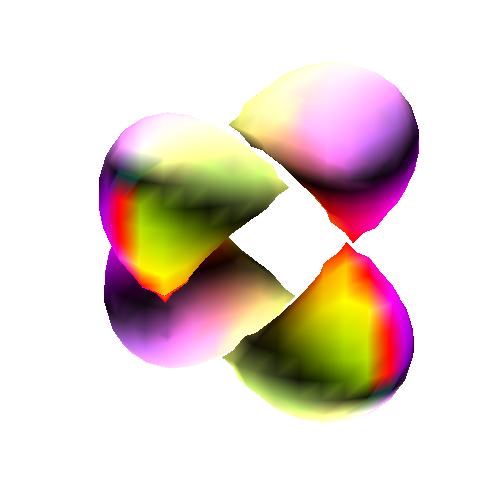}}\\
    \mbox{
      \includegraphics[height=0.0795\textheight]{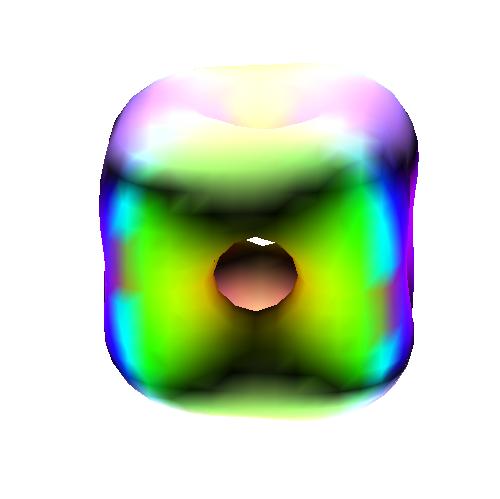}
      \includegraphics[height=0.0795\textheight]{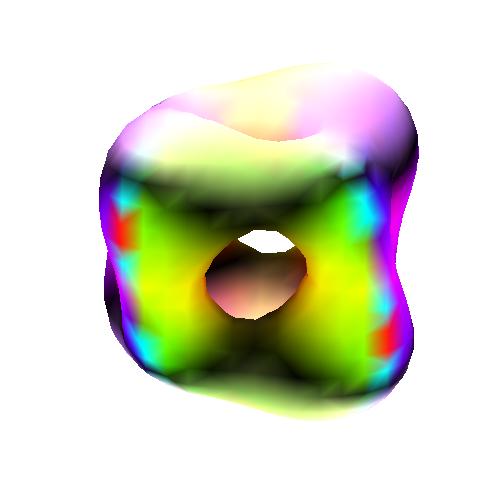}
      \includegraphics[height=0.0795\textheight]{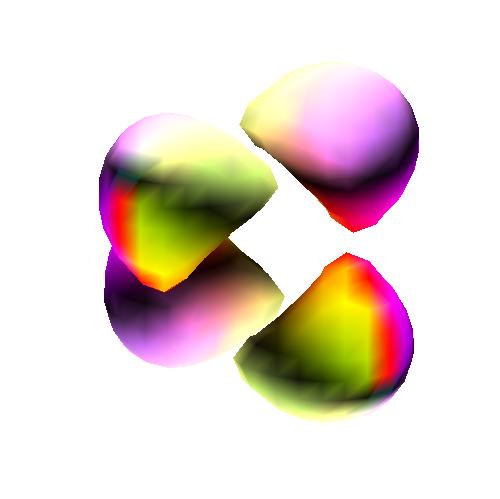}}\\
    \mbox{
      \includegraphics[height=0.0795\textheight]{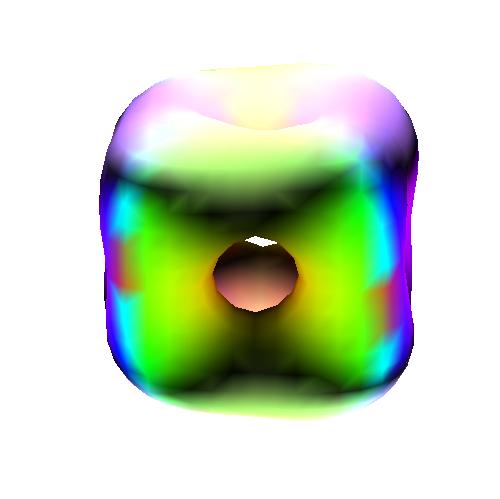}
      \includegraphics[height=0.0795\textheight]{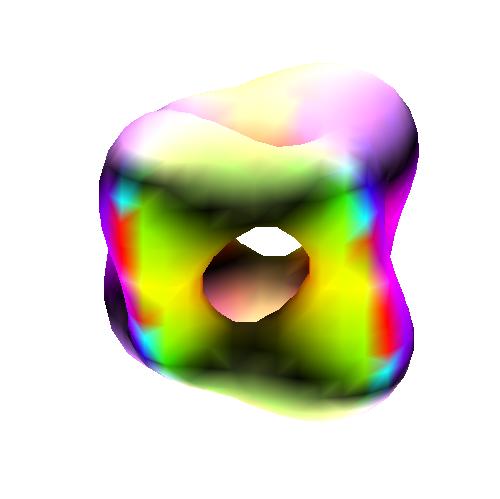}
      \includegraphics[height=0.0795\textheight]{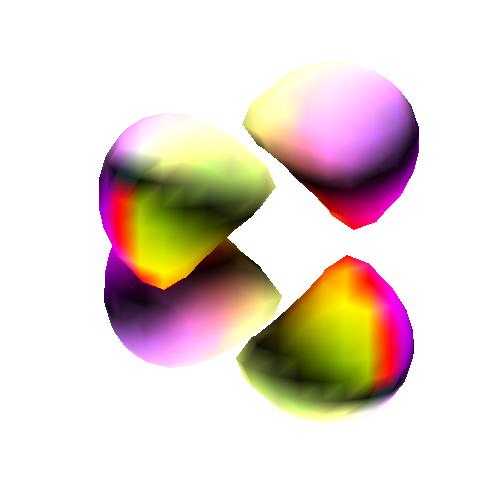}}\\
    \mbox{
      \includegraphics[height=0.0795\textheight]{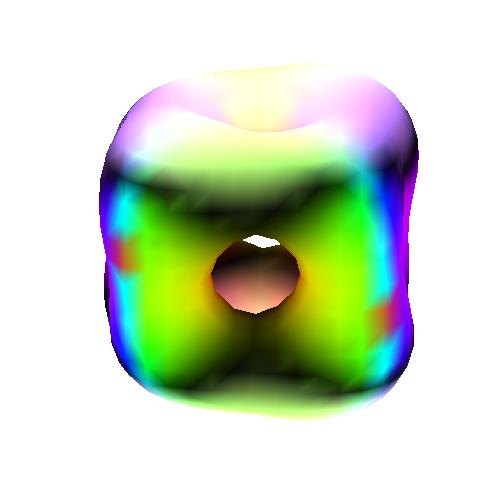}
      \includegraphics[height=0.0795\textheight]{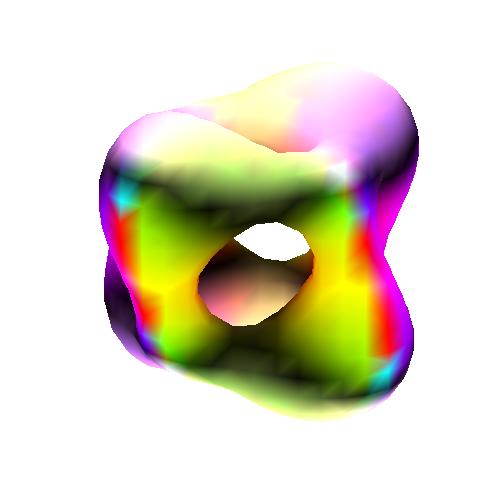}
      \includegraphics[height=0.0795\textheight]{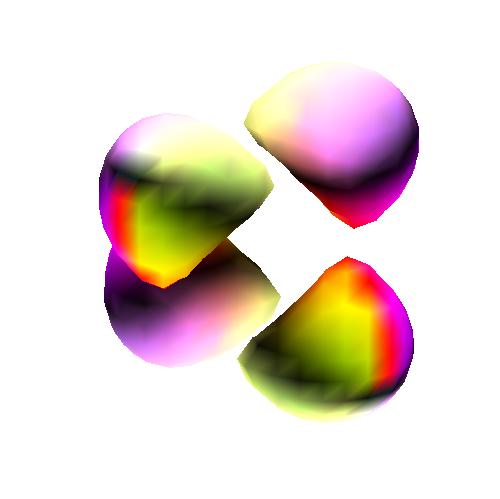}}\\
    \mbox{
      \includegraphics[height=0.0795\textheight]{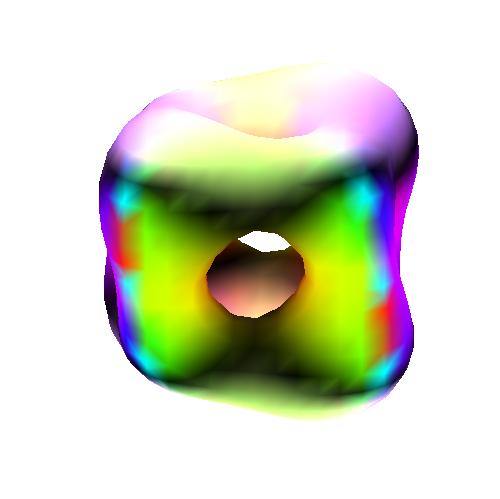}
      \includegraphics[height=0.0795\textheight]{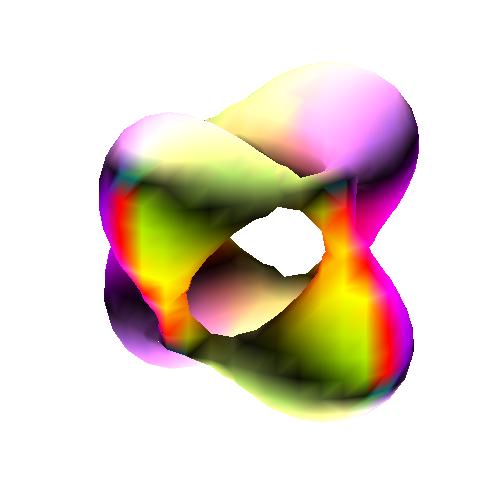}
      \includegraphics[height=0.0795\textheight]{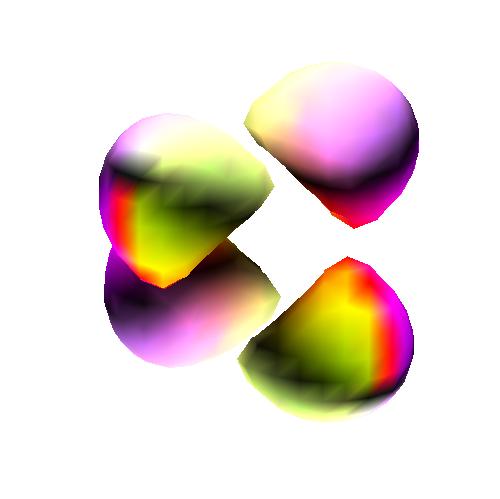}}
    \caption{The columns show $m_\pi=0.375$ Skyrmion solutions with
      $m_2=0.9,1,1.1$ and the rows correspond to $\alpha$ from 0 to
      1 in steps of 0.2 from top to bottom.}
    \label{fig:nout0375}
  \end{minipage}\ \ \ 
  \begin{minipage}[t]{0.47\linewidth}
    \mbox{
      \includegraphics[height=0.0795\textheight]{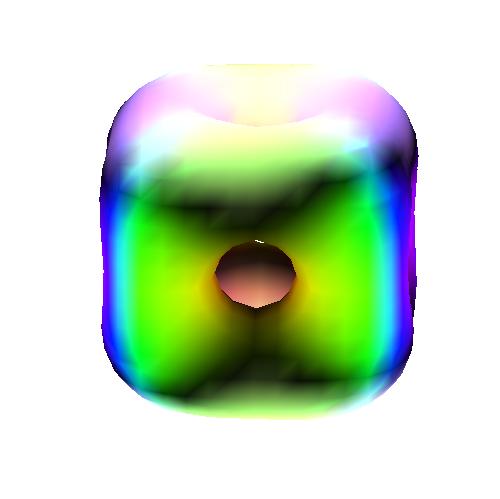}
      \includegraphics[height=0.0795\textheight]{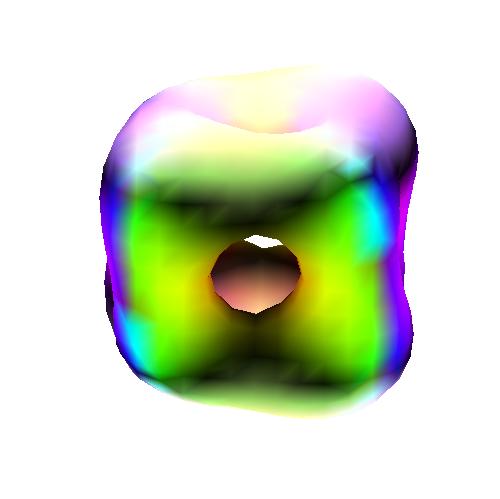}
      \includegraphics[height=0.0795\textheight]{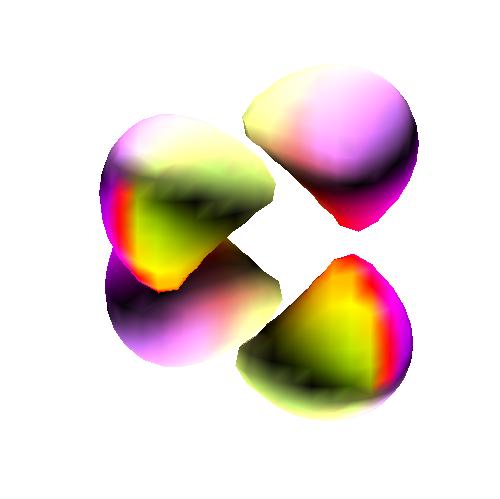}}\\
    \mbox{
      \includegraphics[height=0.0795\textheight]{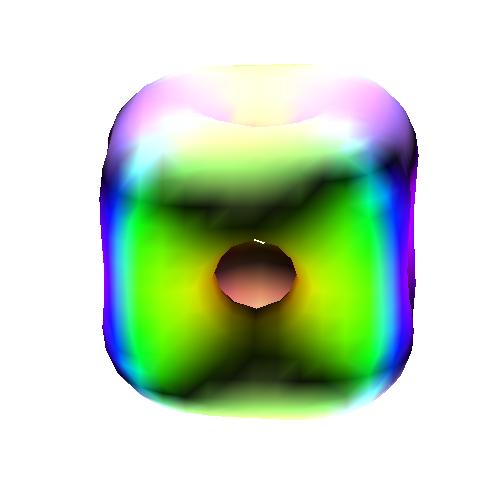}
      \includegraphics[height=0.0795\textheight]{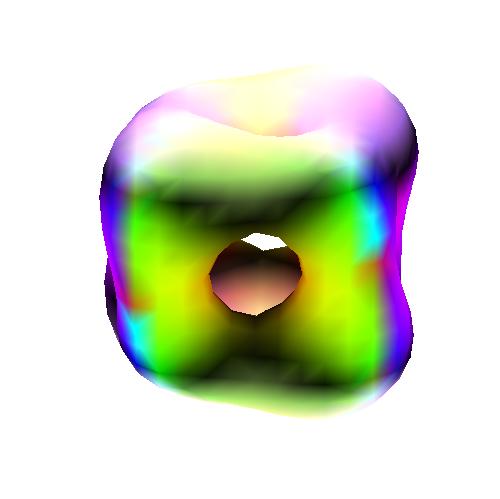}
      \includegraphics[height=0.0795\textheight]{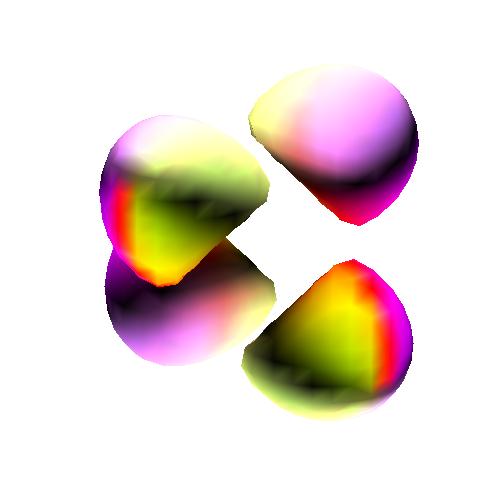}}\\
    \mbox{
      \includegraphics[height=0.0795\textheight]{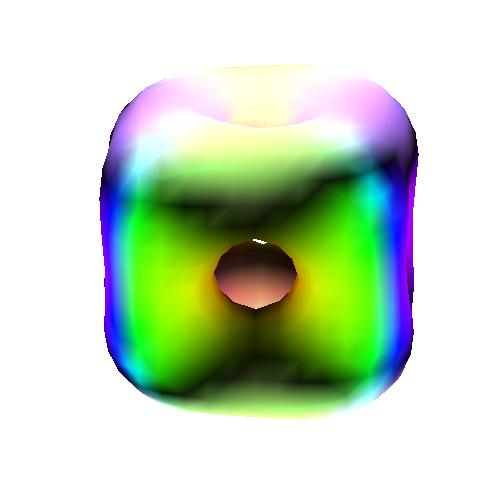}
      \includegraphics[height=0.0795\textheight]{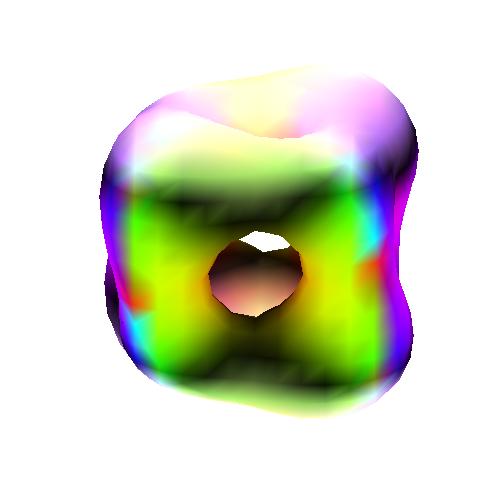}
      \includegraphics[height=0.0795\textheight]{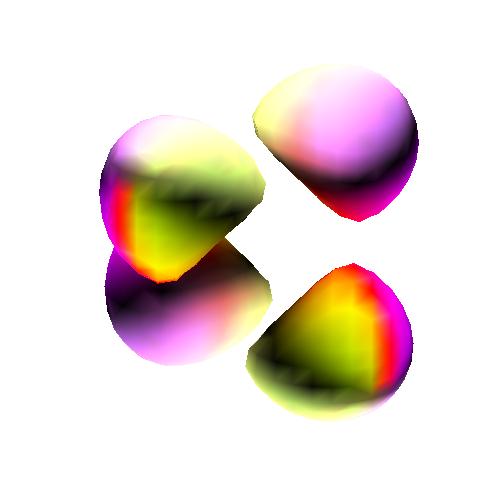}}\\
    \mbox{
      \includegraphics[height=0.0795\textheight]{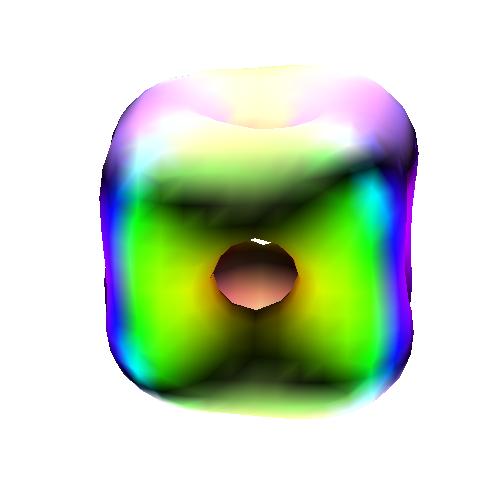}
      \includegraphics[height=0.0795\textheight]{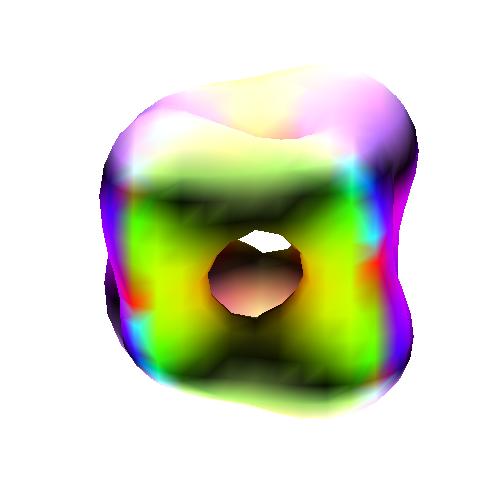}
      \includegraphics[height=0.0795\textheight]{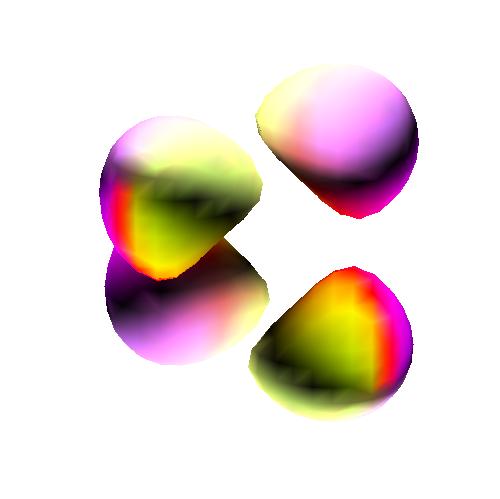}}\\
    \mbox{
      \includegraphics[height=0.0795\textheight]{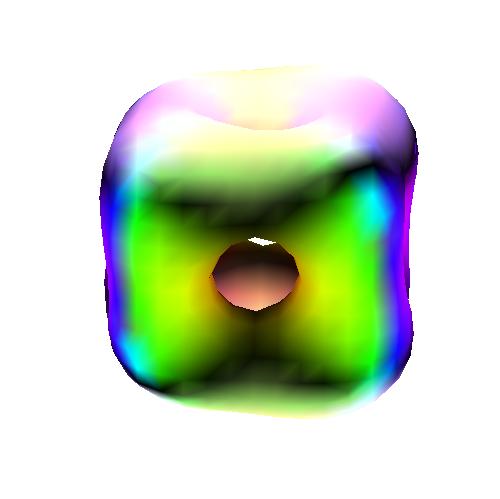}
      \includegraphics[height=0.0795\textheight]{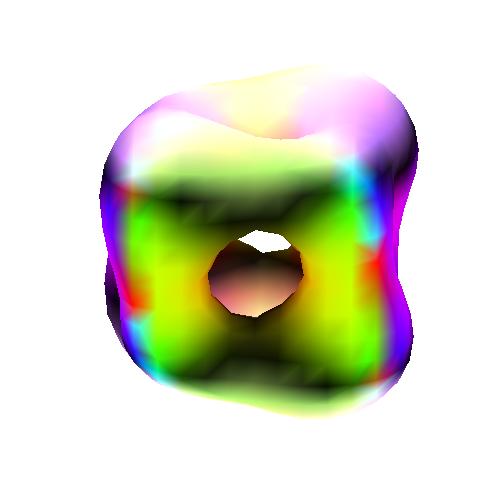}
      \includegraphics[height=0.0795\textheight]{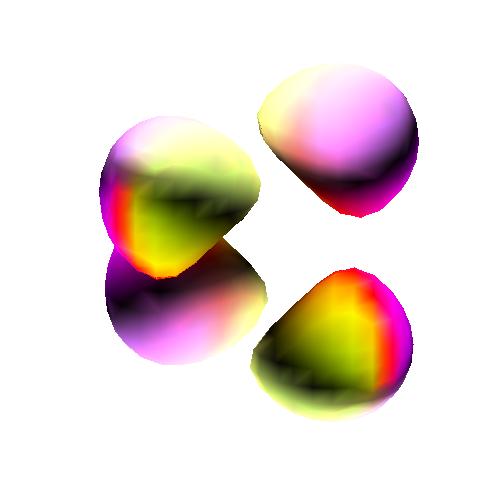}}\\
    \mbox{
      \includegraphics[height=0.0795\textheight]{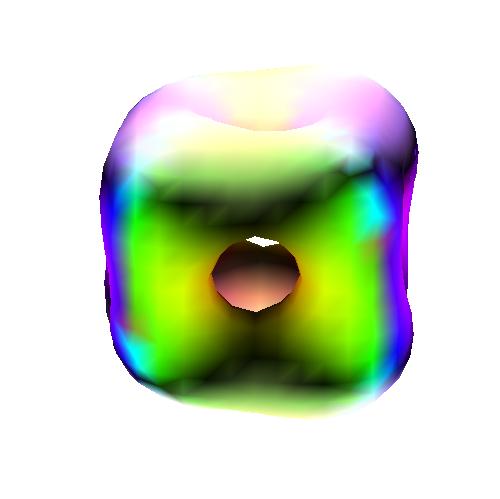}
      \includegraphics[height=0.0795\textheight]{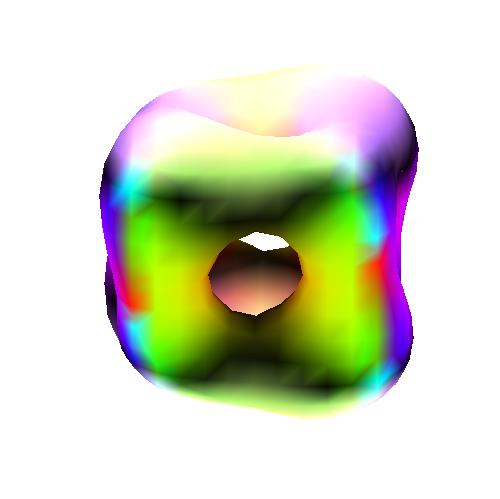}
      \includegraphics[height=0.0795\textheight]{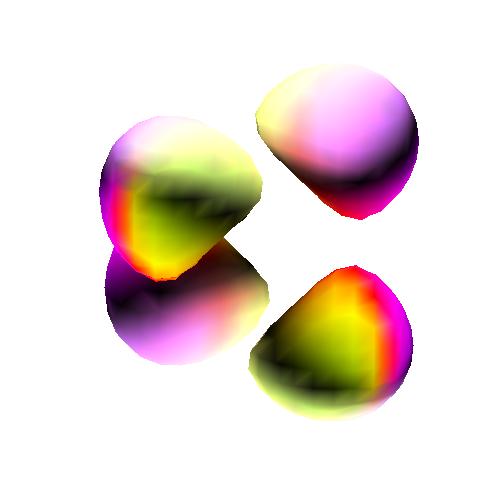}}
    \caption{The columns show $m_\pi=0.5$ Skyrmion solutions with
      $m_2=1.1,1.2,1.3$ and the rows correspond to $\alpha$ from 0 to
      1 in steps of 0.2 from top to bottom.}
    \label{fig:nout05}
  \end{minipage}\ \ 
\end{figure}

Figs.~\ref{fig:nout0125}, \ref{fig:nout025}, \ref{fig:nout0375} and
\ref{fig:nout05} show series of Skyrmion solutions near the boundary
of the mentioned phase transition for $m_\pi=0.125,0.25,0.375$ and
$0.5$, respectively. The loosely bound potential parameter $m_2$ is
increased from left to right in each figure and $\alpha$ is varied
vertically. 

It is interesting to note that the Skyrmions are slightly more
strongly bound and less aloof for $\alpha=0$ (modified pion mass) than
for $\alpha=1$ (standard pion mass).
Consistently with findings in the text, we see that the larger $m_\pi$
is, the larger values of $m_2$ are possible before the phase
transition takes place.

\end{document}